\documentclass[11pt]{article}

\pdfoutput=1 

\usepackage[textwidth = 460 pt, textheight = 650 pt]{geometry}
\usepackage[table]{xcolor}
\usepackage{hyperref,amsthm,amsmath,amssymb,slashed,graphicx,bbold,fancybox,mathtools,color,tikz-cd,caption,subcaption,geometry,marginnote,xcolor,verbatim, marginnote, stmaryrd, manfnt} 
\usepackage{wrapfig}
\usepackage{mathrsfs}
\usepackage{simplewick}
\usepackage{cite, mcite}
\usepackage{microtype}
\usepackage{chapterbib}
\usepackage{cancel}

\usepackage{tikz-cd,tikz}
\usetikzlibrary{arrows,decorations.markings} 

\usepackage[tikz]{mdframed}

\newenvironment{itembox}[1]{\begin{mdframed}[roundcorner=10pt,
  frametitlefont=\normalfont,
  frametitleaboveskip=\dimexpr-0.7\baselineskip,skipabove=\topskip,
  innertopmargin=\dimexpr-0.65\baselineskip,
  innerbottommargin=\dimexpr0.65\baselineskip,
  frametitle={\tikz{\node[anchor=base,rectangle,fill=white] {\strut #1};}}]
}{\end{mdframed}}

\newcommand{\Comment}[1]{{}}

\Comment{\usepackage{xcolor}
\definecolor{MyDarkBlue}{rgb}{0.15,0.15,0.45}
\usepackage[linktocpage=true]{hyperref}
\usepackage{makeidx}
}

\usepackage{framed}
\usetikzlibrary{mindmap,backgrounds}

\definecolor{MyDarkBlue}{rgb}{0.15,0.15,0.45}
\definecolor{shadecolor}{rgb}{0.85,0.85,0.85}
\definecolor{link-blue}{rgb}{0.15,0.15,0.65}
\definecolor{link-red}{rgb}{0.8,0.15,0.1}
\definecolor{link-green}{rgb}{0.15,0.50,0.15}
\definecolor{link}{rgb}{0.45,0.18,0.22}

\usepackage{graphicx}

\def\bpm{\begin{pmatrix}}
\def\epm{\end{pmatrix}}

\def\red{\textcolor{red}}

\def\purple{\textcolor{purple}}
\def\violet{\textcolor{violet}}

\def\magenta{\textcolor{magenta}}

\newcommand{\be}{\begin{equation}}
\newcommand{\ee}{\end{equation}}
\newcommand{\bse}{\begin{subequations}}
\newcommand{\ese}{\end{subequations}}
\newcommand{\bea}{\begin{eqnarray}}
\newcommand{\eea}{\end{eqnarray}}
\newcommand{\dd}{\mathrm{d}}
\newcommand{\nn}{\nonumber}

\hypersetup{
pdftitle={nonlocal charges},
pdfsubject={High Energy Physics},
pdfauthor={Thiago Araujo},
pdfkeywords={gauge; susy; strings; fields; cft},
pdfsubject={hep-th},
colorlinks=true,linkcolor=link,citecolor=link,urlcolor=link,linktocpage
}

\begin{document}

\renewcommand{\thefootnote}{\fnsymbol{footnote}}

\makeatletter
\@addtoreset{equation}{section}
\makeatother
\renewcommand{\theequation}{\thesection.\arabic{equation}}

\rightline{}
\rightline{}

\vspace{0.3cm}

\begin{center}
{\Large \bf{Nonlocal charges from marginal deformations of 2D CFTs:}} \\
\vspace{0.5cm}
{\large \emph{- Holographic \(T \bar T\) \& \(T \bar J\)  and Yang-Baxter deformations -}}
\end{center}
 \vspace{0.7cm}
 
\thispagestyle{empty} 

\centerline{
 {\large { \bf Thiago Araujo \footnote{email:~\href{thgr.araujo@gmail.com}{thgr.araujo@gmail.com}}
}}}
\vspace{0.7cm}

\centerline{{\it
Albert Einstein Center for Fundamental Physics,}}
\centerline{{\it
Institute for Theoretical Physics, University of Bern}}
\centerline{{\it
Sidlerstrasse 5, ch-3012, Bern, Switzerland}}

\vspace{2.0cm}

\thispagestyle{empty}

\centerline{\bf Abstract}

\begin{center}
\begin{minipage}[c]{380pt}
{\noindent  
In this paper we study generic features of nonlocal charges obtained from marginal deformations of WZNW models. Using free-fields representations of CFTs based on simply laced Lie algebras, one can use simple arguments to build the nonlocal charges; but for more general Lie algebras these methods are not strong enough to be generally used. We propose a brute force calculation where the nonlocality is associated to a new Lie algebra valued field, and from this prescription we impose several constraints on the algebra of nonlocal charges. Possible applications for Yang-Baxter and holographic \(T\bar{T}\) and \(T\bar{J}\) deformations are also discussed.
}
\end{minipage}
\end{center}

\vspace{0.3cm}

\begin{flushright}
\emph{\today}
\end{flushright}

\setcounter{page}{1}
\setcounter{tocdepth}{2}

\renewcommand{\thefootnote}{\arabic{footnote}}
\setcounter{footnote}{0}

\newpage

\tableofcontents

\newpage

\section{Introduction and statement of the problem}

\begin{center}
\emph{Marginal deformations}
\end{center}

Integrable deformations of conformal theories form a powerful method to probe physics near a fixed point theory. When the deformation is exactly marginal, we still have, by definition, the constraints imposed by conformal symmetries. When one can additionally consider supersymmetry, gauge symmetry and so on, the final deformed theory can still be manageable from the computational viewpoint. On top of that, integrability may also be taken into account, and in that case it is possible to compute the S-matrix, correlation functions and other observables with a certain degree of generality, although the integrable deformed models are not, obviously, completely tamed.

Deformations of the boundary CFT\(_{d-1}\) correspond, from the AdS/CFT correspondence viewpoint \cite{Maldacena:1997re}, to deformations of the dual AdS\(_d\times {\cal M}_{10-d}\) string solution. When the CFT modification is defined by an exactly marginal operator, the AdS part of the string background remains unchanged, which means that only the internal manifold \({\cal M}_{d-10}\) suffers modifications, as expected. On the other hand, it might happen that the boundary CFTs are deformed by (marginally) relevant or irrelevant operators, and it implies that the deformations necessarily modify the AdS\(_d\) space.

We are obviously interested in all these situations. Exactly marginal deformation is a necessary tool to classify CFTs and understand the geometrical and physical aspects of the moduli spaces. In two dimensions (2D), it is the first step towards a complete classification of the string theory solutions which may ultimately describe the Universe we live in. But we also need to consider the possibility that marginal deformations also receive quantum corrections and break the conformal symmetries. Therefore it is mandatory to develop methods to study the resulting symmetries when we leave the fixed point theory, and it means that we need to consider relevant deformations of CFTs. The resulting massive theories are quite generally harder than the undeformed models, but one might hope for some progress towards their understanding if we assume integrable deformations of the CFTs.

One remarkable example of an exactly marginal perturbation in four dimensions is given by the Leigh-Strassler deformations of \({\cal N}=4\) Super Yang-Mills \cite{Leigh:1995ep}, which give conformal theories with \({\cal N}=1\) supersymmetry. Within the AdS/CFT context \cite{Maldacena:1997re}, the four-dimensional Leigh-Strassler theories provide examples of CFTs with corresponding AdS\(_5\times {\cal M}_5\) dual solutions obtained as deformations of the AdS\(_5\times \mathbb{S}^5\) solution. The gravity dual to the Leigh-Strassler theories are not known in general, but for real \(\beta\)-deformations the dual supergravity solution has been built through sequence of a T-duality followed by a shift, and a second T-duality (TsT-transformation) of the AdS\(_5\times \mathbb{S}^5\) solution, while for complex \(\beta\)-deformations we use either an \(SL(3,\mathbb{R})\) transformation or an STsTS  \cite{Lunin:2005jy}; see also \cite{Aharony:2002hx, Frolov:2005dj}. 

Moreover, the Lax pairs of the TsT-transformed solutions have been found in \cite{Frolov:2005dj}, and these objects settle the classical integrability of the deformed models. Interestingly enough, the TsT transformation is just one example of a bigger and more powerful set of integrable deformations, namely the so-called Yang-Baxter deformations \cite{Klimcik:2002zj, Delduc:2013fga, Delduc:2013qra, Kawaguchi:2014qwa, Matsumoto:2015jja}. Associating these new integrable backgrounds to their elusive holographic dual, one can learn a great deal about the surprising mathematical and physical aspects of a vast collection of quantum field theories, see \cite{Seiberg:1999vs, Hashimoto:1999ut, Maldacena:1999mh, Matsumoto:2014nra, Matsumoto:2014gwa, Matsumoto:2014ubv, Matsumoto:2015uja, Kawaguchi:2014fca, vanTongeren:2015uha, Arutyunov:2015mqj, Wulff:2016tju, Borsato:2016ose, Kyono:2016jqy, Orlando:2016qqu, Matsumoto:2015ypa, Osten:2016dvf, Borsato:2016pas, vanTongeren:2016eeb, Sakamoto:2017cpu, Sakamoto:2017wor, Araujo:2017jkb, Araujo:2017jap, Araujo:2017enj, Fernandez-Melgarejo:2017oyu, Araujo:2018rbc, Fernandez-Melgarejo:2018wpg, Lust:2018jsx, Sakamoto:2018krs, Orlando:2018kms, Orlando:2018qaq, Catal-Ozer:2019tmm, Orlando:2019rjg} for a nonexhaustive list of references. 

\begin{center}
	\emph{Nonlocal charges from marginal deformations}
\end{center}

Two-dimensional theories are of great interest for string theorists and statistical physicists. Fortunately enough, conformal symmetry and integrability in two dimensions are much more restrictive thanks to holomorphicity properties. As one important example of the power of two-dimensional theories for our present work, there is a theorem which determines the necessary and sufficient conditions for exactly marginality of 2D CFT deformations given by a product of holomorphic and antiholomorphic currents, namely \({\cal O}_{(1,1)}=J(z)\bar{J}(\bar{z})\) \cite{Chaudhuri:1988qb, Borsato:2018spz}; see also \cite{Orlando:2006cc}. Additionally, it has been argued that this class of marginal deformations can be written as \(O(d,d)\) transformations  \cite{Hassan:1992gi, Henningson:1992rn, Kiritsis:1993ju, Forste:1994wp, Forste:2003km, Israel:2003ry, Orlando:2006cc, Catal-Ozer:2019tmm, Orlando:2019rjg}. In this respect, the authors \cite{Orlando:2019rjg} extended important results on integrable exactly marginal deformations of two-dimensional CFTs using \(O(d,d)\) deformations, and they have also shown how to build the nonlocal Lax pairs of the deformed models which guarantee their classical integrability.

We should remark that integrable theories are often defined in terms of commuting local conserved charges, but for some special models it is known that integrability can be equivalently formulated in terms of noncommuting local and nonlocal charges, and it is the most important concept behind the \emph{Yangians}. As for the very definition of integrability, there exist good reasons to study the presence of nonlocal charges in local field theories, but it is definitely not limited to this aspect. In other words, nonlocal charges also have a prominent role in condensed matter theories, in anyonic and parafermionic statistics, superselection sectors and in the axiomatization of quantum field theories. As we see in the text, one may think of these nonlocal objects as analogues of the 't Hooft operators; consequently, a better terminology for these nonlocal charges would be disorder operators. 

The analogy with the 't Hooft operator becomes even more conspicuous with the observation by Ricci \emph{et. al.} \cite{Ricci:2007eq} that has shown how local (Noether) charges and nonlocal charges are exchanged by T-duality. This result is similar to the well-known fact that Wilson loops and 't Hooft operators are exchanged by Hodge duality. Notwithstanding the limitation of this simple analogy, this idea underpins the current work at the most fundamental level: We want to find the algebra of a collection of disorder operators obtained from deformation. Additionally, as the authors of  \cite{Orlando:2019rjg} pointed out, nonlocality of the Lax pair should not be considered as a surprise since these charges induced by the nonlocal Lax pairs may be thought of as a result of the stringy aspects of their approach. In other words, it should be considered as a benefit rather than a drawback of this analysis. 

\begin{center}
	\emph{Motivation}
\end{center}

The scrutiny of these nonlocal charges cannot be thought of as a simple refinement of quantum field theories. In the way we understand these objects nowadays, they suddenly appear in different corners of physics, but there is no complete understanding of their general properties whatsoever. On the other hand, the argument that we are deepening our comprehension of field theories could be used to justify a very broad variety of research lines. In fact, the present paper was driven by a series of pragmatic questions. First of all, as we said before, the authors of \cite{Ricci:2007eq} have shown how T-duality exchanges local and nonlocal charges. Although their analysis was inspired by one specific scenario, the original AdS\(_5\)/CFT\(_4\) duality, one may ask if it is possible to devise a situation where all local charges, and consequently their algebras, are dualized (or deformed) to nonlocal charges with a consistent corresponding algebra.

 Furthermore, classical nonlocal charges have also appeared in the Yang-Baxter literature, see for example Frolov \cite{Frolov:2005dj} in his analysis of the integrability aspects of the Maldacena-Lunin background. It has also been suggested that the resulting symmetries of Yang-Baxter deformed models are obtained as Drinfel'd twists \cite{Matsumoto:2014cja, Matsumoto:2015jja, vanTongeren:2015uha, Araujo:2017jkb, Araujo:2017jap, Loebbert:2018lxk, Dlamini:2019zuk} of the original symmetries. Hence, one may also expect the same twists extended to the Yangian structure of the \({\cal N}=4\) SYM \cite{Dolan:2004ps, Beisert:2010jq, Torrielli:2010kq, Spill:2012qe, Loebbert:2016cdm}. In other words, when we have an integrable system with known Yangian symmetry, one may naturally expect that the resulting deformed model is also endowed with a Yangian obtained from the twist of the original one.
 
These two aspects show how our methods to build nonlocal charges strongly rely on the integrability of models. Quite remarkably, Matsumoto and Yoshida \cite{Matsumoto:2014cja, Matsumoto:2015jja} may teach us how we can leave the integrable systems realm. They have also shown that the deformed currents can be written as nonlocal gauge transformations of local currents. One important aspect of this particular result is that it closes the question on the existence of these nonlocal objects after Yang-Baxter deformations: they are quite ubiquitous. Consequently, once again we may ask ourselves if it is possible to consider a model (not necessarily integrable) so severely deformed that all global conserved charges are converted into nonlocal ones. 

\begin{center}
	\emph{Summary of the paper}
\end{center}

Therefore we have some distinct but complementary problems summarized as follows. Given a generic, not necessarily integrable, deformation of the CFT, we would like to know if, and how, one can define the nonlocal charges of the deformed theory in terms of the undeformed local ones. This babe problem is summarized in the case (b) in the table below. Observe that knowing how the nonlocal charges of the original theory change under deformation is an important step if we want to know how the Yangians are deformed; for that reason it would be of great interest to understand all of the following four maps.
\[
\begin{tabular}{lll}
	\hline \hline
	(a) \texttt{local} \(\mapsto\) \texttt{local}& & \red{(b) \texttt{local} \(\mapsto\) \texttt{nonlocal}} \\ 
	& & \\
	(c) \texttt{nonlocal} \(\overset{?}{\mapsto}\) \texttt{local} & & (d) \texttt{nonlocal} \(\overset{?}{\mapsto}\) \texttt{nonlocal} \\ 
	\hline \hline
\end{tabular} 
\]

 As for the existence of these charges, inspired by the ideas of Bernard and LeClair \cite{Bernard:1990ys}, briefly described in section 2.3, we propose a very economical approach to define nonlocality in  section 3.1 via free field representations of the CFTs. As we see, an important consequence of this construction is that the CFT Hilbert space structure is preserved. Sadly enough, if the symmetry group is sufficiently large the calculations become cumbersome very fast, and it is obviously an unavoidable drawback of this approach. 

Inspired by the same methods of \cite{Bernard:1990ys} and by the deformations via nonlocal gauge transformations of Matsumoto and Yoshida \cite{Matsumoto:2014cja, Matsumoto:2015jja}, we try a daring proposal to describe the nonlocal currents in section 3.2. We conjecture the existence of a \(\widehat{\mathfrak{g}}_k\)-algebra valued field which defines the massive theory obtained by deformations. The CFT limit of this new field is given by the sum of left and right currents which define the Kac-Moody algebras \(\hat{\mathfrak{g}}_L\times \hat{\mathfrak{g}}_R\). More specifically, the field can be written as \({\cal A}={\cal L}+\bar{{\cal R}}\) and satisfies our beloved free-fields equation of motion \(\partial \bar{\partial}{\cal A}=0\) at the fixed point. In other words, \({\cal A}\) is fairly trivial as an object in the CFT, but it may become a fundamental dynamical object in the deformed theory, and understanding its role for nonlocal charges is part of our proposal in section 3.2.

Additionally, this second brute force proposal is much more general and allows us to understand the Hilbert space structure of the deformed theory. As a matter of fact, the Hilbert space decomposition of the CFT is not generally preserved under relevant deformations, and imposing this decomposition, see for example in \cite{Zamolodchikov:1989zs}, seems to be a bold conjecture in the current state of affairs. From the technical point of view, it means that the deformed current operator product expansions (OPEs) have nontrivial structures and that the usual decomposition between holomorphic and antiholomorphic sectors does not exist if the deformation is not exactly marginal. In section 4 we start developing some methods to study these OPEs, and the algebras generated by the charges.

In section 5 we try to see how these ideas may be applied (in spite of the current limitations) to Yang-Baxter deformation, and in the \(T\bar{T}\) and \(T\bar{J}\) deformations \cite{Zamolodchikov:2004ce, Caselle:2013dra, Smirnov:2016lqw, Cavaglia:2016oda, Guica:2017lia}. As we said before, the ideas concerning Yang-Baxter deformations are one of the reasons behind the present paper, but the use of this technology in a different context may appear as a good surprise. First of all, at the leading order, the operator \(T\bar{T}\) is built from the holomorphic and antiholomorphic terms of the stress-energy tensor, while \(T\bar{J}\) comes from the holomorphic part of the energy-momentum tensor, but also needs a \(U(1)\) current \(\bar{J}(\bar{z})\). Consequently, these deformations are irrelevant from the renormalization group (RG) viewpoint with conformal dimensions \((2,2)\) and \((2,1)\), respectively, and many interesting aspects of these transformations have been addressed, for example, \cite{Donnelly:2018bef, Conti:2018jho, Conti:2018tca, Bonelli:2018kik, Baggio:2018rpv,  Conti:2019dxg, Banerjee:2019ewu, Sfondrini:2019smd, Chang:2019kiu, Jiang:2019hux}.

Using the AdS/CFT correspondence in our favor, one can easily notice that given an irrelevant deformation of the boundary (or holographic) CFT, the string background should suffer a deformation as well. Evidently, the strings propagating in this new deformed background are described by a worldsheet CFT. Consequently, the counterpart of the irrelevant deformation of the holographic CFT must be marginal from the worldsheet perspective. In this respect, the authors of \cite{Giveon:2017nie, Giveon:2017myj, Chakraborty:2018vja, Apolo:2018qpq, Chakraborty:2019mdf, Giveon:2019fgr} have shown how we can construct certain exactly marginal deformations of the worldsheet CFT that describe the most important features -- rather than all the features -- of the \(T\bar{T}\) and \(T\bar{J}\) deformations of the boundary CFT. Unfortunately, the methods we use here do not seem to be strong enough to be applied within the context of exactly marginal deformations, but we still can try to understand some aspects of these deformations.

It is well documented that nonlocality is also associated to nontrivial braiding relations and disorder operators, and that it is also important in the study of anyons and parafermions \cite{Fateev:1985mm, Fredenhagen:1988fj, Frohlich:1988vt, Lerda1992, Loebbert:2016cdm}. The message we want to leave is that nonlocal structures are poorly understood even in the easiest cases where they are present. In this paper we address some problems, and propose some methods to study them.

\section{Nonlocal charges for WZNW-models}

As we have just explained, the construction of nonlocal charges is a difficult problem to handle, and in order to study them in full generality, it is necessary to develop new techniques first. Evidently there are well-known results that can be used as guiding principles. This section intends to be a review of these well-known results, and I also hope to bring together some calculations scattered in the literature.

\subsection{Nonlocal charges}

We start with the action for a \(\sigma\)-model based on the Lie algebras \(\mathfrak{g}\) given by
\bse
\be
S = S_0 + S_{WZ}\; ,
\ee
where we have the respective terms
\be
S_0 = \frac{k}{4\pi}\int_{\partial {\cal B}} \dd^2 \sigma \mathrm{Tr}\left(\partial^\mu g^{-1}\partial_\mu g \right)
\; ,\quad
S_{WZ}=
-\frac{\alpha k}{6\pi} \int_{\cal B}\dd^3\tilde{\sigma}\epsilon^{\alpha\beta\gamma}\mathrm{Tr}\left(  g^{-1}\partial_\alpha g  g^{-1}\partial_\beta g  g^{-1}\partial_\gamma g \right)\; 
\ee
\ese
with \(\alpha\) being the coupling constant and k the level of the theory. When the running constant reaches the value \(\alpha=1\) we have the conformal fixed point described by the WZNW (Wess-Zumino-Novikov-Witten) model. Once we consider the current
\be 
j_\mu = g^{-1}\partial_\mu g\; ,
\ee 
and the condition \(\partial_\mu g^{-1}=-g^{-1}\partial_\mu g g^{-1}\), the equations of motion can be written as 
\be 
\partial_\mu {\cal R}^\mu\equiv \partial_\mu j^\mu + \alpha \epsilon^{\mu\nu} \partial_\mu j_\nu = 0\; ,
\ee
with the conventions \( \epsilon^{01}=-\epsilon^{10}=1\). Similarly, we could consider the fields \(\tilde{j}_\mu = \partial_\mu g g^{-1}\) with corresponding currents \({\cal L}_\mu=\tilde{j}_\mu - \alpha \epsilon_{\mu\nu}\tilde{j}^\nu\).

At the risk of repeating ourselves, the relevant information about these models is carried by the conserved currents
\be 
\begin{split}
{\cal R}_\mu(t,x) & = j_\mu (t,x) +\alpha \epsilon_{\mu\nu}j^\nu(t,x)\\
{\cal L}_\mu(t,x) & = \tilde{j}_\mu (t,x) -\alpha \epsilon_{\mu\nu}\tilde{j}^\nu(t,x)\; ,
\end{split}
\ee
where \(\alpha\) is a running constant whose value \(\alpha=1\) gives a fixed point in the RG flow. It has been known for quite some time that these are the basic ingredients to construct the nonlocal currents of the theory \cite{deVega:1979zy, Abdalla:1984gm, Bernard:1991vq, Bernard:1996bp, Abdalla:1991vua, MacKay:1992he}. Let us write them as
\bse
\be 
\label{nonlocal1}
\mathcal{K}_\mu(t,x) = \overbrace{(1-\alpha^2)\epsilon_{\mu\nu} j^\nu(t,x)}^{\equiv\; \mathcal{K}^{(0)}} +
\overbrace{\frac{1}{2} [{\cal R}_\mu(t,y),\int_{{\cal C}_x} \star{\cal R}(t',y)]}^{\equiv\; \mathcal{K}_\mu^{(1)}}
\ee
and 
\be 
\label{nonlocal2}
\mathcal{I}_\mu(t,x) =\overbrace{ (1-\alpha^2)\epsilon_{\mu\nu}\tilde{j}^\nu(t,x)}^{\equiv \; \mathcal{I}_\mu^{(0)} } +\overbrace{\frac{1}{2} [{\cal L}_\mu(t,x),\int_{{\cal C}_x} \star{\cal L}(t',y)]}^{\equiv\;  \mathcal{I}_\mu^{(1)} }\; ,
\ee
\ese
where the nonlocality is governed by the line integral along any curve \({\cal C}_x\in \mathbb{R}^2\) connecting the points \(\{ (t,-\infty) ,(t,x) \}\), and it is parametrized by \((t',y)\). Observe that in the CFT limit only the commutators survive.

Although not strictly necessary for our future considerations, let us for completeness verify the conservation of the nonlocal currents (\ref{nonlocal1}) and (\ref{nonlocal2}). Using that \(\star {\cal R}= -\epsilon_{\mu\nu}  {\cal R}^{\mu} \dd x^\nu \), we have
\be 
\partial^\mu \mathcal{K}_\mu(t,x) = (1-\alpha^2)\epsilon^{\mu\nu} \partial_{\mu}j_\nu(t,x)
+\frac{1}{2} [{\cal R}^{\mu}(t,x),\int_{{\cal C}_x} \partial_\mu \star{\cal R}(t',y)]\; ,
\ee
where we have used that \(\partial_\mu{\cal R}^{\mu}=0 \). We can now calculate the two pieces separately. The first term is
\be 
\partial^\mu \mathcal{K}_\mu^{(0)}=(1-\alpha^2)\epsilon^{\mu\nu} \partial_{\mu}j_\nu = (1-\alpha^2)(\partial_0 j_1 - \partial_1 j_0) = -(1-\alpha^2)[j_0, j_1]\; ,
\ee
where we have used the identity \(\partial_0 j_1-\partial_1 j_0 + [j_0, j_1]=0\), while the second term gives
\be 
\partial^\mu \mathcal{K}_\mu^{(1)}= 2 \left(1-\alpha^2 \right) [j_0,j_1]\; .
\ee
Putting all these facts together we see that \(\partial^\mu \mathcal{K}_\mu(t,x)=0\), which means that the nonlocal current (\ref{nonlocal1}) is conserved. Evidently one can repeat the analysis for (\ref{nonlocal2})  to show that \(\partial^\mu \mathcal{I}_\mu(t,x)=0\). The important aspect of this analysis is that these conservation laws are independent on the value of the running coupling \(\alpha\), therefore, these objects can be defined for any nonlinear sigma model, and not only in the fixed point \(\alpha=1\).

At the conformal point we can write the currents in a particularly useful form. Let us start seeing that for \(\alpha=1\) we have
\bse
\be 
{\cal R}_\mu = j_\mu+ \epsilon_{\mu\nu}j^\nu\; ,
\ee
which essentially means that 
\be 
{\cal R} \equiv {\cal R}_0 = - {\cal R}_1 \qquad \textrm{where}\qquad 
\left\{  
\begin{array}{l}
{\cal R}_0= j_0 - j_1\\
{\cal R}_1= - j_0 + j_1
\end{array}
\right.\; .
\ee
\ese
Using the light-cone coordinates \(z=t+ x\) and \(\bar{z}=t - x\), one can show that the above current satisfies
\be 
\partial \bar{{\cal R}} = 0\; \; ,\quad \text{where}\quad \bar{{\cal R}} = g^{-1}\bar{ \partial} g \; .
\ee
This equation clearly implies the antiholomorphicity \( \bar{\cal R}= \bar{\cal R}(\bar{z}) \). Moreover, one can easily show that 
\be 
\bar \partial {\cal R} = -[\bar {\cal R}, {\cal R}]\; ,
\ee
and it readily implies the flatness condition 
\be 
\bar \partial {\cal R}  - \partial \bar{{\cal R}}
 + [\bar {\cal R}, {\cal R}] =0\; .
\ee
On similar grounds we have \({\cal L}=\partial g g^{-1}\), and that implies the relation \(\bar{\partial}{\cal L}=0\) and the holomorphicity \({\cal L}={\cal L}(z)\). As it is well known, these currents define the Kac-Moody algebras which emerge from the OPEs
\bse
\begin{align}
{\cal L}^a(z){\cal L}^b(w)\sim & \frac{k \delta^{ab}}{(z-w)^2}+\frac{i f^{ab}_{\phantom{ab}c}}{z-w}{\cal L}^c(w)\label{ope1}\\
\bar{\cal R}^a(\bar z)\bar{\cal R}^b(\bar w)\sim & \frac{k \delta^{ab}}{(\bar z-\bar w)^2}+\frac{i f^{ab}_{\phantom{ab}c}}{\bar z-\bar w}\bar{\cal R}^c(\bar{w})\label{ope2}\; .
\end{align}
\ese

Once we define the nonlocal charges  (\ref{nonlocal1}) and (\ref{nonlocal2}), we need to find a good prescription to describe their quantum counterparts. Using the results of \cite{Luscher:1977uq, Abdalla:1984gm, Bernard:1990ys, Bernard:1991vq}, a useful prescription is given by the point-splitting regularization
\be 
\label{split}
\begin{split}
[j_\mu(x+\delta), j_\nu(x)] & = C_{\mu\nu}^\rho(\delta)j_\rho(x) + D_{\mu\nu}^{\rho\lambda}(\delta)\partial_\rho j_\lambda(x)\\
[\tilde{j}_\mu(x+\delta), \tilde{j}_\nu(x)] & = \widetilde{C}_{\mu\nu}^\rho(\delta)\tilde{j}_\rho(x) + \widetilde{D}_{\mu\nu}^{\rho\lambda}(\delta)\partial_\rho \tilde{j}_\lambda(x)\; ,
\end{split}
\ee
where the OPE coefficients \(C\) and \(D\) can be obtained from the conservation laws and additional physical constraints, such as the CPT theorem and locality \cite{Luscher:1977uq, Abdalla:1984gm, Bernard:1990ys, Bernard:1991vq}. Roughly speaking, in the next section we see that in order to define nonlocal conserved currents for the deformed models, we need to solve a system of nonlinear coupled differential equations, which is difficult to solve. On the other hand, we can obtain some intuition on the algebra defined by these currents using an expansion inspired by (\ref{split}) and imposing the conservation law and appropriate physical conditions.

At any rate, it is well known that the quantized currents are defined as 
\be 
\begin{split}
\mathcal{K}_\mu(t,x) &   =\lim_{\delta\to 0_+} \left((1-\alpha^2)Z(\delta)\epsilon_{\mu\nu} j^\nu(t,x) +\frac{1}{2} [{\cal R}_\mu(t,x),\int_{{\cal C}_x} \star{\cal R}(t',y|\delta)]\right)\\
\mathcal{I}_\mu(t,x) &  =\lim_{\delta\to 0_+}\left( (1-\alpha^2)\epsilon_{\mu\nu}\widetilde{Z}(\delta)\tilde{j}^\nu(t,x) +\frac{1}{2} [{\cal L}_\mu(t,x),\int_{{\cal C}_x} \star{\cal L}(t',y|\delta)]\right)\; ,
\end{split}
\ee
where \({\cal R}(t',y|\delta)\) and \({\cal L}(t',y|\delta)\) are written in terms of the point-splitting regularization (\ref{split}). Additionally, the coefficients \(Z(\delta)\) and \(\widetilde{Z}(\delta)\) are multiplicative counterterms. 

In conclusion, we have the following situations:
\begin{itemize}
	\item In a principal chiral model point, \(\alpha=0\), we can easily build our nonlocal charges and therefore, our Yangian symmetry when the Wess-Zumino term goes to \(0\) \cite{Loebbert:2016cdm}.
	\item It is known that there are no anomalies in the renormalized currents \cite{Abdalla:1984gm, Bernard:1991vq}, and that the corresponding quantum nonlocal charges are preserved in a very straightforward manner. These objects generate the Yangian symmetry \({\cal Y}_L(G)\times {\cal Y}_R(G)\).
	\item The perturbed theory is evidently harder to be analyzed, but there is at least one situation where the analysis is straightforward, namely, when the deformed theory is described by a sigma model. In other words, if the deformed theory is another sigma model based on the Lie algebra \(\mathfrak{h}\subseteq \mathfrak{g} \) and \(\alpha\neq 1\), the analysis we performed above is easily applied. This situation is, as one may imagine, very unique and demands a combination of factors.
\end{itemize}

In the case where we do not know if the deformed theory is described by a sigma model, the method above is useless and we need to develop other techniques to build our nonlocal charges.

\subsection{Marginally relevant deformations}

We want to consider perturbations of conformal field theory of the form
\be 
\label{pert-cft}
{\cal S} = {\cal S}_{CFT} + \eta \int \dd^2 x{\cal O}_{(\Delta, \bar{\Delta})}(x)\; ,
\ee
where \({\cal O}_{(\Delta, \bar{\Delta})}(x)\) is an operator of conformal dimension \((\Delta, \bar{\Delta})\). When our CFT is defined in the IR region, one may consider that the perturbation operator above is relevant, \((\Delta<1, \bar\Delta<1)\), which essentially means that it is a small perturbation in the UV limit, but becomes large in the infrared region. For spinless fields we have \(\Delta=\bar \Delta\), and it naturally implies that the coupling \(\eta\) has conformal dimensions \((1-\Delta, 1-\Delta)\). 

Remember that we are in a quantum theory; therefore the expressions below are defined inside correlation functions, and it ultimately means that everything is properly renormalized \cite{Cardy:1989da}. It is important to keep this point in mind since we consider perturbations which are classically marginal but that can have anomalous dimensions. These are the marginally relevant deformations that play a fundamental role in the present work. In summary, our perturbations are of the form
\bse
\be 
\label{perturb}
\delta {\cal S}= \eta\int \dd^2x {\cal O}_{(1,1)}(x)\; ,
\ee 
or more specifically, we consider deformations generated by a current-current operator
\be 
\label{jj.pert}
\delta {\cal S}= \eta\int \dd^2x (j_{(1,0)}\bar{j}_{(0,1)})(x)\; .
\ee 
\ese
In the case of marginally relevant deformations, the coupling \(\eta\) receives quantum corrections and has dimensions
\be 
[\eta]\simeq (\delta(\eta_0), \delta(\eta_0))\; ,
\ee
where \(\eta_0\) is the bare coupling. This deformation is marginally relevant provided \(0<\delta<1\), and in that particular case we can try to understand the perturbed theory using the well established techniques of Zamolodchikov, Bernard and LeClair \cite{Zamolodchikov:1989zs, Bernard:1990ys}. 

The hallmark of two-dimensional conformal field theories is the existence of a traceless and conserved stress-energy tensor, 
\be 
T_{z\bar z}=0\; , \quad \bar\partial T(z)=0 \; , \quad \partial \bar T(\bar z)=0\; .
\ee
Moreover, factorization of (unitary) CFTs implies that any field \(T^{(k)}\)  of conformal weight \((h,0)\), where \(k\) just enumerates these fields, satisfies the holomorphicity condition
\be 
\bar \partial T^{(k)}=0\; .
\ee
Once we perturb the theory, we should not expect these two distinctive properties unless the perturbation is given by an exactly (or truly) marginal operator. Quite generally, we have a relation of the form
\be 
\bar \partial T^{(k)}(z, \bar{z}) = \partial \Theta^{(k)}(z, \bar z) \equiv \lambda R^{(k)}\; ,
\ee
where the stress-energy tensor occurs for, say, \(k=1\), and the nondiagonal element of the symmetric stress-energy tensor is \(\Theta^{(1)}(z , \bar z)\). Moreover, we have assumed that only terms of order \(\mathscr{O}(\lambda)\) contribute to this equation. For further details, including the notation, we refer to the original paper \cite{Zamolodchikov:1989zs}.

Zamolodchikov has also shown that given perturbations of the form (\ref{perturb}), the corrections for relevant deformations are given by the equations
\be 
\label{zam.eq}
\begin{split}
\bar{\partial} T^{(k)}_{s+1}(z, \bar{z}) & = \frac{\eta}{2\pi i}\oint \dd w {\cal O}(w,\bar{z}) T^{(k)}_{s+1}(z)\\
\partial \bar{T}^{(k)}_{s+1}(z, \bar{z}) & = \frac{\eta}{2\pi i}\oint \dd \bar{w} {\cal O}(z,\bar{w})\bar{T}^{(k)}_{s+1}(\bar{z})\; .
\end{split}
\ee
where \( {\cal O} \equiv {\cal O}_{(1,1)}\) and \(s\) is an additional index that labels the spin; therefore, the stress-energy tensor would correspond to \((k,s)=(1,1)\). We refer to these formulas as the \emph{Zamolodchikov's equations}, and they play a fundamental role in our discussion.

\subsection{Sine-Gordon and SU(2) WZNW at level k=1}

The construction of nonlocal charges from marginal deformations of WZNW models has been partly  addressed in \cite{Bernard:1990ys} from a very different point of view. Consider the sine-Gordon model
\be 
S = \frac{1}{4\pi}\int \dd^2 z\left( 
\partial \Phi \bar{\partial}\Phi + 4 \lambda : \cos(\hat{\beta} \Phi):
 \right)\; ,
\ee
where the interaction term \(4 \lambda : \cos(\hat{\beta} \Phi)\) is treated as a perturbation of the free-field theory. When the parameter assumes the value \(\hat{\beta}=\sqrt{2}\), the perturbation is marginal and it can be described as a current-current deformation of the \(SU(2)_1\) WZNW model. Therefore, the sine-Gordon model can be described as 
\be 
S = S^{k=1}_{SU(2)} + \frac{\lambda}{2\pi}\int \dd^2 z \left({\cal L}^+ \bar{\cal R}^- + {\cal L}^- \bar{\cal R}^+ + g {\cal L}^0 \bar{\cal R}^0   \right)\; ,
\ee
where \(g\) breaks the \(SU(2)_1\) symmetry defined by the OPEs
\be 
\begin{split} 
{\cal L}^0(z) {\cal L}^0(w) & \sim \frac{1}{2(z-w)^2}\\
{\cal L}^0(z){\cal L}^\pm(w) & \sim \pm \frac{{\cal L}^\pm(w)}{(z-w)}\\
{\cal L}^+(z){\cal L}^-(w) & \sim \pm \frac{1}{(z-w)^2} +  \frac{2 {\cal L}^0(w)}{(z-w)}\; ,
\end{split}
\ee
and similarly for the right-moving currents \(\bar{ \cal R}\).

For the free boson theory, it is well known that we can write the field \(\Phi(z,\bar{z})\) as
\be 
\Phi(z,\bar{z}) = \phi(z) + \bar{ \phi}(\bar{z})\; , \qquad \langle  \phi(z) \phi(w)\rangle = - \ln (z-w)\; ,
\ee
with primary fields \(V_\alpha(z) =: e^{i \alpha \phi(z)}:\) of weight \(\Delta = \alpha^2/2\). Evidently, it is easy to see that \(V_{\pm \sqrt{2}}(z)\equiv {\cal L}^\pm(z)\) have conformal weight \(1\). Moreover, the vertex operators \(V_{\alpha}(z)\) satisfy nontrivial braiding relations, and using this fact Bernard and LeClair studied the quantum group structure of the perturbed theory \cite{Bernard:1990ys}. In the free boson representation we have
\be 
{\cal L}^0(z)=\frac{i}{\sqrt 2} \partial \phi(z)\; , \qquad {\cal L}^\pm(z) = :e^{\pm i \sqrt{2}\phi(z)}:\; .
\ee

\begin{figure}[ht]
\begin{center}
\includegraphics[scale=0.45]{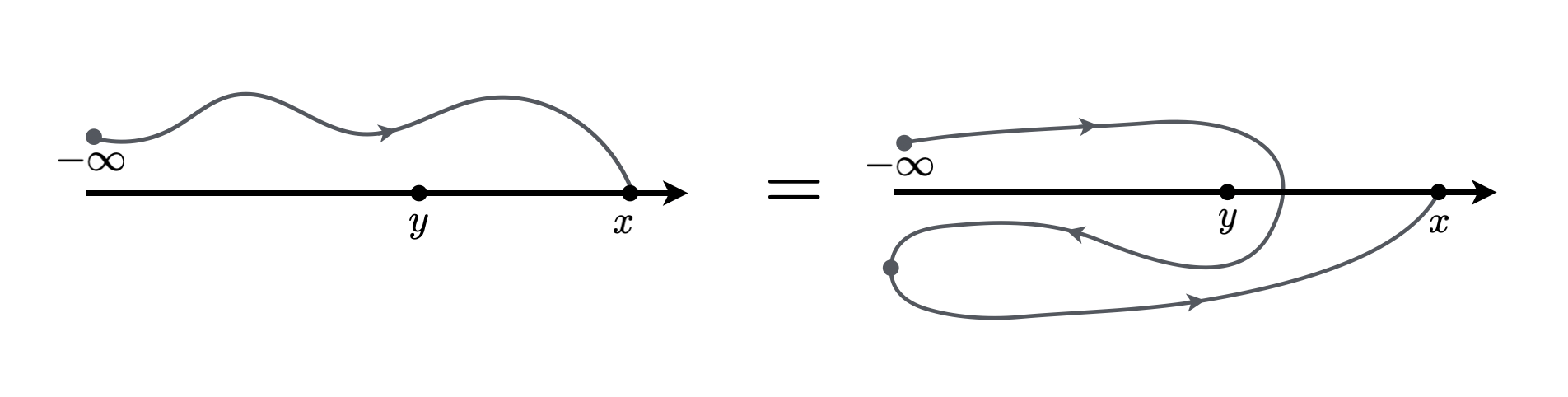}	
\caption{Topological Obstruction}
\label{topobs}
\end{center}
\end{figure}

The field \(\phi(t,x)\) is no longer holomorphic once we perturb the theory, but in the special case where the decomposition of \(\Phi(x,t)\) is preserved, we can write
\be 
\label{nlfields}
\phi(t,x) = \frac{1}{2}\left( \Phi(x,t) + \int_{-\infty}^x \dd y \partial_t \Phi(y,t) \right)\quad
\bar{\phi}(t,x) = \frac{1}{2}\left( \Phi(x,t) - \int_{-\infty}^x \dd y \partial_t \Phi(y,t) \right)
\ee
where these formulae can be verified from the conditions \(\bar{\partial}\phi(z)=\partial\bar{\phi}(\bar{z})=0\) that are satisfied at the conformal fixed point. Nonlocality is a consequence of the integral in \((-\infty, x)\) above. The most immediate consequence of this condition is the existence of equal-time braiding relations
\be 
J_\mu^a(t,x) J_\nu^b(t,y) = \mathrm{R}^{ab}_{cd} J_\nu^c(t,x) J_\mu^d(t,y) \ \ y<x \; .
\ee
Moreover, one can easily verify that the matrix \(\mathrm{R}\) satisfies the quantum Yang-Baxter equation. One can think of these braidings as topological obstructions when we move a disorder operator as in the figure \ref{topobs}.
As a matter of fact, the terminology disorder operator is much more appropriate for the objects we are considering, and it would avoid possible confusions that the term nonlocal could bring, especially when we are in fact studying a local quantum field theory. In any case, we continue using the standard terminology.

Nonlocal currents in the sine-Gordon theory, the deformed \(SU(2)_1\) model, are \(({\cal L}^\pm(z), {\cal H}^\pm(z))\), where \({\cal L}^\pm\) are now defined in terms of the fields (\ref{nlfields}). The explicit expressions for \({\cal H}^\pm\) have been written in \cite{Bernard:1990ys}, and it is immediate to verify that \({\cal H}^\pm(z)\overset{\lambda \to 0}{=}0\). These currents satisfy \(\bar{\partial} {\cal L}^\pm = \partial {\cal H}^\pm\) and there are, evidently, the corresponding expressions for the antiholomorphic sector, namely \(\partial \bar{\cal R}^\pm = \bar{\partial}\bar{\cal H}^\pm\). The nonlocal charges are
\be 
Q^\pm = \frac{1}{2\pi i }\left( \int \dd z {\cal L}^\pm + \int \dd \bar{z} {\cal H}^\pm \right)\; \quad 
\bar{Q}^\pm = \frac{1}{2\pi i }\left( \int \dd \bar{z} \bar{\cal R}^\pm + \int \dd z \bar{\cal H}^\pm \right)\; .
\ee
Putting all these facts together, Bernard and LeClair have shown that the deformed algebra of nonlocal charges is given by \(\widehat{sl(2)}_q\), where the deformation parameter is simply \(q=\exp(-\sqrt{2}\pi i)\). One should also observe that the deformation parameter of the q-deformed algebra \(\widehat{sl(2)}_q\) does not depend on the deformation parameter \(\eta\).

\section{Nonlocal operators from vertex operators}

How could we generalize the methods that were useful in the \(SU(2)_1\) WZNW? Evidently the answer to this problem is not straightforward; otherwise it would already have been answered. We address it in the present section, and in particular, we discuss some ideas about a complete development of new techniques to build and unveil the algebra of nonlocal charges in more general WZNW models.  In the absence of a definite answer, now we discuss two different directions.

\subsection{Nonlocal charges from free-field representations}

Our initial proposal is to use the free-field representation for WZNW models as a starting point of nonlocality. This is a very direct generalization of what has been discussed by Bernard and Le Clair, and consequently the applicability is extremely limited  \cite{Bernard:1990ys}, as we show now. Given a simply laced Lie algebra \(\mathfrak{g}\), the free-field representation says that for all simple roots we have a corresponding free boson \(\phi^i\) with
\be 
\langle \phi^i(z)\phi^j(w)\rangle\sim -\delta^{ij}\ln(z-w)\; ,
\ee
and with currents
\be
H^j(z) = i \partial \phi^j(z)\; , \quad \tilde{E}^\alpha(z) = e^{i\alpha\cdot \phi(z)}\; ,
\ee
where \(\alpha\cdot \phi(z) = \sum_j \alpha^j \phi^j(z)\). For further details see our favorite CFT books \cite{DiFrancesco:1997nk, Ketov:1995yd}. 

From these equations and from the expressions (\ref{nlfields}), we can immediately generalize what Bernard and LeClair have done for the \(su(2)_1\) model in \cite{Bernard:1990ys}. In other words, we have a simple collection of expressions of the form
\be 
\phi^i(t,x) = \frac{1}{2}\left( \Phi^i(x,t) + \int_{-\infty}^x \dd y \partial_t \Phi^i(y,t) \right)\quad
\bar{\phi}^i(t,x) = \frac{1}{2}\left( \Phi^i(x,t) - \int_{-\infty}^x \dd y \partial_t \Phi^i(y,t) \right)\; ,
\ee
one for each simple root. Consequently, we also need to write down a list of OPEs for the currents and for the vertex operators of conformal weight \(\Delta=1\). Furthermore, the explicit expressions for the deformed currents are strongly dependent on the specific perturbation we are interested in, but there is no challenge in writing down the explicit formulas once we fix the operator which defines the perturbation. It would also be a straightforward, but laborious, exercise to determine the deformation that gives the quantum group \(\widehat{sl(n)}_q\) as residual symmetries.

For non simply laced Lie algebras, the presence of short roots makes the situation a bit harder, but it may be possible to apply similar ideas. For instance, operators of the form \(\exp(\pm i \alpha\cdot \phi)\) have conformal dimension \(1/2\), and in order to transform them into currents, we need to multiply them by free fermions. As an example, the WZNW associated to the algebra \(B_n\) demands \(n\) bosons and one additional fermion. In the deformed theory, if the nonlocality is defined just in terms of the bosonic field \(\phi\) as before, leaving the free fermionic field local, the previous analysis is very trivial to be performed. On the other hand, we expect that in a more general situation, both the bosonic and fermionic fields define some sort of nonlocality, and it demands new mathematical tools.

Observe that what we have said up to now can be also extended to Lie algebras at any level \(k>1\). Remember that in those cases, it is also possible to use free-field representations, and we simply need to define the Cartan generators as \(H^j = i\sqrt{k}\partial \phi^j\). The vertex operators \(\exp(\pm i \alpha\cdot \phi/\sqrt{k})\) have conformal dimensions \(\Delta=|\alpha|^2/(2k)\), and as in the non simply laced case, in order to obtain our currents we simply multiply the vertex operators by new fields of appropriate conformal dimensions. Remarkably, these new operators have interesting nonlocal properties and are called parafermions \cite{Fateev:1985mm, DiFrancesco:1997nk, Mussardo:2010mgq}. 

Finally, for general Lie algebras \(\widehat{\mathfrak{g}}_k\) we would need to use a stronger approach, for example, the Wakimoto free-field representation, but the drawbacks of these free-fields inspired approaches are much more evident. For example, once we assume a very generic deformation in this particular representation, the \((\beta,\gamma)\)-system is also be modified and the explicit expressions for the nonlocal charges would be much more involved. On the other hand, we see in section \ref{string} that there are situations where some progress can be made, and that this representation may be useful to determine the nonlocal charges for deformations of the AdS\(_3\times \mathbb{S}^3\times {\cal M}_4\) strings.

Some of these issues are addressed in section \ref{string}, but we hope to fill computational details of this section in a future publication. For now, we recognize that there are severe limitations in the free-fields representation and we  try to find a better and stronger approach. We unfortunately do not provide a final answer to this point, but the next section is a sort of brute-force calculation, and although it may seem a contrived construction at first, it might be possible to learn more about the algebra of nonlocal charges. 

\subsection{Nonlocal charges from deformed WZNW-models currents}

The authors of \cite{Kawaguchi:2013lba, Kawaguchi:2012ug, Matsumoto:2014cja} have shown that in the context of integrable deformations, the deformed currents can be written as nonlocal gauge transformations. More specifically, they have shown that given an integrable field theory with currents \(J^{\pm}=g^{-1}\partial_\pm g\), where \(g\) is the Lie group element, the Yang-Baxter deformed theory has resulting currents given by \(\tilde{J}_\pm = \tilde{g}^{-1}\partial_\pm \tilde{g}\), where \(\tilde{g}=\mathcal{F}^{-1}g\) are nonlocal fields defined through a twist element \(\mathcal{F}\) whose explicit form is in the original papers. Therefore, the nonlocal currents are finally written as 
\bse
\be 
\label{nl.curr}
\tilde{J}_\pm = \mathcal{G}J_\pm \mathcal{G}^{-1} - \partial_\pm \mathcal{G} \mathcal{G}^{-1}\; ,
\ee 
where \(J_\pm\) are the original undeformed currents and \(\mathcal{G}=g^{-1}\mathcal{F}g\). This relation is one of the reasons behind the study of symmetries of the Yang-Baxter deformed theories, and it suggests that the symmetries are obtained through a Drinfel'd twist of the original ones. These ideas have been explored in many texts, including \cite{Matsumoto:2014cja, Matsumoto:2015jja, vanTongeren:2015uha, Araujo:2017jkb, Araujo:2017jap, Loebbert:2018lxk, Dlamini:2019zuk}. 

The form (\ref{nl.curr}) is the finite form of a Yang-Baxter deformation. For an infinitesimal expansion of the field \(\mathcal{G}\simeq 1 + \eta\, (\texttt{nonlocal})\), where \(\eta\) is a perturbative parameter, one could suppose that the infinitesimal version of (\ref{nl.curr}) has the following structure
\be 
\label{inl.curr}
\tilde{J}_\pm = J_\pm \pm\eta\ (\texttt{nonlocal term})\; .
\ee
For the marginal deformations of our interest, we look for nonlocal currents with this structure.
\ese

More specifically, the marginally relevant current-current deformation (\ref{jj.pert}) is written as 
\be 
S = S_{cft} + \frac{\eta}{2\pi} \sum_{a, b} c_{ab} \int\dd^2z   {\cal L}^a(z) \bar{\cal R}^{b}(\bar{z})\; ,
\ee 
where \(\eta\) is an infinitesimal parameter and \(c_{ab}\) is the matrix that defines the specific deformation we are interested in. Furthermore, it is known under which conditions these deformations are exactly marginal \cite{Chaudhuri:1988qb, Orlando:2006cc, Borsato:2018spz}. We should also observe that the deformations above are not, in general, \(\mathfrak{g}\) invariant. Consequently we cannot apply the usual set of ideas \cite{Luscher:1977uq, Bernard:1990jw, Hauer:1997ig, Hauer:1997gw}, and in particular, the massive theories do not define (local) massive current algebras as defined by Bernard in \cite{Bernard:1990jw}.

In what follows, we use different symbols for the fields at the fixed point theory, namely \({\cal L}_\ast^a\) and \(\bar{\cal R}_\ast^a\), and we keep the notation \({\cal L}^a\) and \(\bar{\cal R}^a\) for the perturbed theory. In order to construct an object of the form (\ref{inl.curr}), let us now define the \(\mathfrak{g}\)-valued function
\bse
\be 
\label{field}
\mathcal{A}_\ast(z, \bar{z}) = {\cal L}_\ast(z) + \bar{\cal R}_\ast(\bar{z})\; ,
\ee
in the infrared conformal field theory. This object is very trivial in the CFT point, and it obviously satisfies the free boson equation of motion
\be 
\label{Aeq}
\partial \bar{\partial} \mathcal{A}_\ast(z, \bar{z}) = 0 \; .
\ee
Once we move along the RG flow, one may assume the structure (\ref{field}) is preserved, that is 
\be 
\mathcal{A}(t, x) = {\cal L}(t, x) + \bar{\cal R}(t, x)\; ,
\ee
\ese
and in this sense, one can try to use the same methods of \cite{Bernard:1990ys}. Moreover, using that \(\partial = (\partial_0 + \partial_1)/2\) and \(\bar{\partial} =(\partial_0 - \partial_1)/2\) we find
\be 
\partial \bar{\cal{R}}_\ast = 0 \; \Leftrightarrow \; \partial_0 \bar{\cal{R}}_\ast = - \partial_1 \bar{\cal{R}}_\ast\qquad \text{and} \qquad 
\bar{\partial} {\cal L}_\ast = 0 \; \Leftrightarrow \; \partial_0 {\cal L}_\ast = \partial_1 {\cal L}_\ast\; .
\ee
Therefore, we can write the following objects
\be 
\label{nloc.}
\mathcal{L} = \frac{1}{2}\left( \mathcal{A} + \int_{-\infty}^{x}\dd y \partial_0 {\cal A}(t, y)  \right)\qquad
\bar{\mathcal{R}} = \frac{1}{2}\left( \mathcal{A} - \int_{-\infty}^{x}\dd y \partial_0 {\cal A}(t,y)  \right)\; ,
\ee
which are analogous to the expressions (\ref{nlfields}). Once again, in the fixed point these two equations are identities, but as we move along the RG flow one may consider (\ref{nloc.}) as nonlocal definitions of the conserved currents \({\cal L}\) and \(\bar{{\cal R}}\). 

One should observe that these currents are not necessarily associated to any hidden Yangian. In other words, the deformations above are not necessarily integrable, therefore, these nonlocal currents should not be considered as objects (necessarily) built from the monodromy matrix. In fact, the presence of these fields is not in general clear, but it may be a generalization (in the sense that we do not require integrability of the system) of an exotic symmetry as those studied in \cite{Kawaguchi:2010jg, Kawaguchi:2011pf, Kawaguchi:2011mz, Kawaguchi:2012ug, Kawaguchi:2013gma, Kawaguchi:2013lba}. 

The field \({\cal A}(z, \bar{z})\) does not give any additional information in the CFT limit since it is just a sum of the chiral and antichiral components of the currents, but in the massive theory, this field has its own dynamical equation that is a modification of (\ref{Aeq}). Using the Zamolodchikov's equations (\ref{zam.eq}) we have 
\be
\label{zam.con}
\bar{\partial}\left( \partial {\cal A}\right)(z,\bar{z}) = \frac{\eta}{2\pi i} \oint \dd w {\cal O}(w, \bar{z}) \partial {\cal L}_\ast(z)\; , \quad
\partial \left( \bar{\partial} {\cal A}\right)(z,\bar{z}) = \frac{\eta}{2\pi i} \oint \dd \bar{w} {\cal O}(z, \bar{w}) \bar{\partial}\bar{{\cal R}}_\ast(\bar{z})\; .
\ee
These two conditions should obviously be the same, and this equality enters as an additional constraint. On the other hand, we want to find an expression for the field \({\cal A}\). Using the explicit form of the marginal operator, we see that the first condition gives
\bse
\be
\begin{split} 
\bar{\partial}\left( \partial {\cal A}^a(z,\bar{z})\right) & = \frac{\eta}{2\pi i}c_{a'b'}\oint \dd w {\cal L}_\ast^{a'}(w) \partial {\cal L}_\ast^a(z) \bar{{\cal R}}_\ast^{b'} (\bar{z})\\
& = i c_{a'b'}\eta f^{a'a}_{\phantom{a'a}c} \partial {\cal L}_\ast^c(z) \bar{\cal R}_\ast^{b'}(\bar{z})\\
& = i c_{a'b'}\eta f^{a'a}_{\phantom{a'a}c} \partial {\cal A}_\ast^c(z, \bar{z}) \bar{\cal R}_\ast^{b'}(\bar{z})\; .
\end{split}
\ee
In the last equality we have used the condition \(\partial {\cal L}_\ast^c(z)=\partial {\cal A}_\ast^c(z, \bar{z})\) that obviously holds in the IR fixed point theory. Finally, one may consider the \(\bar{\partial}\)-derivative
\be 
\label{eq.01}
\bar{\partial}^2\left( \partial {\cal A}^a(z,\bar{z})\right) =  i c_{a'b'}\eta f^{a'a}_{\phantom{a'a}c}  \left( \partial\bar{\partial} {\cal A}_\ast^c(z, \bar{z}) \bar{\cal R}_\ast^{b'}(\bar{z}) + \partial {\cal A}_\ast^c(z, \bar{z}) \bar{\partial}{\cal A}_\ast^{b'}(z,\bar{z}) \right)\; .
\ee
\ese
\bse
Moreover, we can repeat the same idea for the second expression in (\ref{zam.con})
\be
\begin{split} 
\partial \left( \bar{\partial} {\cal A}^a(z,\bar{z})\right) & = \frac{\eta}{2\pi i} c_{a'b'}\oint \dd \bar{w} {\cal L}_\ast^{a'}(z) \bar{{\cal R}}_\ast^{b'} (\bar{w}) \bar{\partial} \bar{{\cal R}}_\ast^a(\bar{z})\\
& = i c_{a'b'} \eta  f^{b'a}_{\phantom{b'a}c} {\cal L}_\ast^{a'}(z) \bar{\partial} \bar{\cal R}_\ast^c(\bar{z})\\
& = i c_{a'b'} \eta  f^{b'a}_{\phantom{b'a}c} {\cal L}_\ast^{a'}(z) \bar{\partial} {\cal A}_\ast^c(z,\bar{z})\; .
\end{split}
\ee
Taking the \(\partial\)-derivative
\be 
\label{eq.02}
\partial^2 \left( \bar{\partial} {\cal A}^a(z,\bar{z})\right) = i c_{a'b'} \eta  f^{b'a}_{\phantom{b'a}c}  \left({\cal L}_\ast^{a'}(z) \partial\bar{\partial}{\cal A}_\ast^c(z, \bar{z}) + \partial{\cal A}_\ast^{a'}(z, \bar{z}) \bar{\partial} {\cal A}_\ast^c(z,\bar{z}) \right)\; .
\ee
\ese

Therefore if we impose the equation \(\partial\bar{\partial}{\cal A}_\ast^a=0\) on the rhs of (\ref{eq.01}) and (\ref{eq.02}), their sum and difference yield respectively 
\bse 
\be\label{dyn.eq.a}
\begin{split}
\partial\bar{\partial}\left( \bar{\partial} {\cal A}^a(z,\bar{z}) + \partial{\cal A}^a(z,\bar{z}) \right) & \overset{!}{=} i c_{a'b'}\eta \left( f^{a'a}_{\phantom{a'a}c} \partial {\cal A}^c(z, \bar{z}) \bar{\partial}{\cal A}^{b'}(z,\bar{z}) + f^{b'a}_{\phantom{b'a}c}  \partial{\cal A}^{a'}(z, \bar{z}) \bar{\partial} {\cal A}^c(z,\bar{z}) \right)\\
\partial\bar{\partial}\left( \bar{\partial} {\cal A}^a(z,\bar{z}) - \partial{\cal A}^a(z,\bar{z}) \right) & \overset{!}{=} i c_{a'b'}\eta \left( f^{a'a}_{\phantom{a'a}c} \partial {\cal A}^c(z, \bar{z}) \bar{\partial}{\cal A}^{b'}(z,\bar{z}) - f^{b'a}_{\phantom{b'a}c}  \partial{\cal A}^{a'}(z, \bar{z}) \bar{\partial} {\cal A}^c(z,\bar{z}) \right)\; ,
\end{split}
\ee 
where we considered that \({\cal A}\simeq {\cal A}_\ast+{\cal O}(\eta)\) to make the replacement \({\cal A}_\ast\mapsto {\cal A}\). These equations are rather involved, but we may notice that we could initially try to find a solution for \(X^a=\partial{\cal A}^a\) and \(Y^a=\bar{\partial}{\cal A}^a\), so that the equations above become 
\be\boxed{
\begin{aligned}\label{dyn.eq.b}
\partial\bar{\partial}\left( X^a(z,\bar{z}) + Y^a(z,\bar{z}) \right) & = i c_{a'b'}\eta \left( f^{a'a}_{\phantom{a'a}c} X^c(z, \bar{z}) Y^{b'}(z,\bar{z}) + f^{b'a}_{\phantom{b'a}c}  X^{a'}(z, \bar{z}) Y^c(z,\bar{z}) \right)\\
\partial\bar{\partial}\left( X^a(z,\bar{z}) - Y^a(z,\bar{z}) \right) & = - i c_{a'b'}\eta \left( f^{a'a}_{\phantom{a'a}c} X^c(z, \bar{z}) Y^{b'}(z,\bar{z}) - f^{b'a}_{\phantom{b'a}c}  X^{a'}(z, \bar{z}) Y^c(z,\bar{z}) \right)\; ,
\end{aligned}}
\ee
\ese 
and now we have a system of 2 \(\dim(\mathfrak{g})\) second order differential equations with unknowns \(X^a\) and \(Y^a\), \(a=1,\dots, \dim(\mathfrak{g})\). In other words, this system of nonlinear differential equations defines the field \(\mathcal{A}\), which is the necessary object to define the nonlocal charges of the deformed WZNW model. 

\subsubsection*{Currents conservation}

In the deformed theory, we also need to verify if the nonlocal currents are associated to a new form of symmetry in the system, possibly similar to those exotic symmetries as in \cite{Kawaguchi:2010jg, Kawaguchi:2011pf, Kawaguchi:2011mz, Kawaguchi:2012ug, Kawaguchi:2013gma, Kawaguchi:2013lba}. Our first test is to certify that the nonlocal currents define conservation laws. Using Zamolodchikov's equations (\ref{zam.eq}) we find
\be
\label{cons.laws01}
\begin{split} 
\bar{\partial}{\cal L}^a(z, \bar{z}) & = \frac{\eta}{2\pi i} c_{a'b'} \oint_z\dd w \left( {\cal L}_\ast^{a'}(w) {\cal L}_\ast^a(z)\right) \bar{\cal R}_\ast^{b'}(\bar{z})= \eta c_{a'b'} \left( k \delta^{a'a} \partial \bar{{\cal R}}_\ast^{b'}(\bar{z}) + f^{a'a}_{\phantom{a'a}c} {\cal L}_\ast^c(z)  \bar{{\cal R}}_\ast^{b'}(\bar{z}) \right)\\
\partial\bar{\cal R}^a(z, \bar{z}) & = \frac{\eta}{2\pi i}c_{a'b'}\oint_{\bar{z}} \dd \bar{w} {\cal L}_\ast^{a'}(z) \left( \bar{\cal R}_\ast^{b'}(\bar{w})  \bar{{\cal R}}_\ast^a(\bar{z})\right) = \eta c_{a'b'} \left( k \delta^{b'a} \bar{\partial} {\cal L}_\ast^{a'}(z) + f^{b'a}_{\phantom{ab'}c} {\cal L}_\ast^{a'}(z)  \bar{{\cal R}}_\ast^c(\bar{z}) \right) \; ,
\end{split}
\ee
where we have used the OPEs (\ref{ope1}) and (\ref{ope2}). Using that in the fixed point theory we have \(\partial \bar{\cal R}_\ast=0\) and \(\bar{\partial}{\cal L}_\ast=0\), we can write the conditions above as conservation laws
\bse 
\be 
\label{cons.laws02}
\partial_0{\cal L}^a(t,x) = \partial_1 {\cal H}^a(t,x)\; , \quad 
\partial_0 \bar{\cal R}^a(t,x) = \partial_1{\cal K}^a(t,x)\; ,
\ee
where we have defined the following components,
\be 
\begin{split}
\mathcal{H}^a(t,x) & = {\cal L}^a(t,x) +\eta c_{a'b'} f^{a'a}_{\phantom{a'a}c} \int_{-\infty}^x\dd y{ \cal L}_\ast^c(t,y) \bar{\cal R}_\ast^{b'}(t,y)\\
\mathcal{K}^a(t,x)& = - \bar{{\cal R}}^a(t,x) +\eta c_{a'b'} f^{b'a}_{\phantom{b'a}c}\int_{-\infty}^x \dd y {\cal L}_\ast^{a'}(t,y)  \bar{\cal R}_\ast^c(t,y)\; .
\end{split}
\ee
\ese
Together, these conditions show that the nonlocal currents are, indeed, conserved. \footnote{Bernard and LeClair have shown in \cite{Bernard:1990ys} that when the algebra is compact, semisimple and is at the level 1, we have
\[ 
\bar{\partial}{\cal L}^a + \partial\bar{\cal R}^a=0\; .
\]
In this case, we also have 
\[ 
\bar{\partial}{\cal L}^a - \partial\bar{\cal R}^a + 2\eta f^{abc} {\cal L}^b \bar{\cal R}^c =0
\]
and if we define the new structure constants as \(\tilde{f}_\eta^{abc}\equiv \eta f^{abc}\) we have the flatness condition.}. One may also define the following \(\mathfrak{g}\)-valued 1-forms
\bse
\be 
\label{current01}
\begin{split}
J(t,x) & = \left(J^a_0(t,x) \dd t + J^a_{1}(t,x) \dd x\right) T^a\\
& \equiv \left({\cal L}^a(t,x) \dd t + {\cal H}^a(t,x) \dd x \right) T^a\; ,
\end{split}
\ee
and 
\be 
\label{current02}
\begin{split}
\tilde{J}(t,x) & = \left(\tilde{J}^a_0(t,x) \dd t + \tilde{J}^a_{1}(t,x) \dd x\right) T^a\\
& \equiv \left(\bar{{\cal R}}^a (t,x) \dd t + {\cal K}^a (t,x) \dd x \right) T^a\; .
\end{split}
\ee
\ese
Finally, one can conveniently recast the conservation laws (\ref{cons.laws02}) as
\be 
\partial^\mu J_\mu (t,x) = 0\qquad \partial^\mu \tilde{J}_\mu (t,x) = 0\; .
\ee
Moreover, we assume that \(J_\mu(t,x)\),  \(\tilde{J}_\mu(t,x)\), and  \({\cal A}(t,x)\) are the only \(\mathfrak{g}\)-valued fields in the massive theory. This condition means that these objects and the equation of motion for \({\cal A}(z, \bar{z})\) carry all necessary information to define the deformed currents.

It is convenient to write the conservation laws (\ref{cons.laws02}) as
\bse
\begin{align} 
\label{cons.laws03ha}
\bar{\partial}{\cal L}^a(z,\bar{z}) & =\partial \left(\eta c_{a'\bar{b}'} f^{a'a}_{\phantom{a'a}c} h^c(z)\bar{\cal R}^{b'}_\ast(\bar{z})\right)\equiv \partial {\cal H}^a\\
\label{cons.laws03ka}
\partial\bar{\cal R}^{\bar{a}}(z,\bar{z}) & =\bar{\partial} \left(\eta c_{a'\bar{b}'} f^{\bar{b}'\bar{a}}_{\phantom{b'a}\bar{c}} {\cal L}^{a'}_\ast(z)  \bar{h}^{\bar{c}}(\bar{z})\right)\equiv \bar{\partial} {\cal K}^{\bar{a}}\; ,
\end{align}
\ese
where we have used the notation \(a, \bar{a}=1,2, \cdots, \dim(\mathfrak{g})\), which denotes the left- and right-moving symmetries. Hence, from the conserved currents
\bse
\be 
J^a \equiv  (J_z^a, J_{\bar{z}}^a) = ({\cal L}^a, {\cal H}^a)\; \quad 
\tilde{J}^{\bar{a}} \equiv  (\tilde{J}_z^{\bar{a}}, \tilde{J}_{\bar{z}}^{\bar{a}}) = ({\cal K}^{\bar{a}}, \bar{{\cal R}}^{\bar{a}})\; ,
\ee
we can finally define the charges
\be
\label{charges}
\begin{split}
{\cal Q}^{1a} & = \frac{1}{2\pi i}\left( \int \dd z {\cal L}^a(z,\bar{z})+\int  \dd \bar{z} {\cal H}^a(z,\bar{z}) \right)\\ 
{\cal Q}^{2\bar{a}} & = \frac{1}{2\pi i}\left( \int \dd z {\cal K}^{\bar{a}}(z,\bar{z})+\int  \dd \bar{z} \bar{\cal R}^{\bar{a}}(z,\bar{z}) \right)\; .
\end{split}
\ee
\ese

Since the mathematical formalism behind nonlocal (or disorder as it may be much more appropriate) operators in a local quantum field theory is not completely unveiled, we assume, as a general hypothesis, that the symmetry corresponding to these fields mimics the main features of ordinary symmetries associated to local operators. Observe that it is not an \emph{ad doc} construction, but it has been verified in a number of examples related to Yangians and more exotic symmetries, but the vagueness of this hypothesis does not lead us too far. More specifically, we modify some early ideas of  \cite{Luscher:1977uq, Bernard:1990jw, Hauer:1997ig, Hauer:1997gw}, and we assume a broader set of hypotheses.

\begin{itembox}{\textbf{Hypotheses}}
\vspace{0.5cm}
\emph{We assume that the perturbed theory has a nonlocal symmetry with associated \(1\)-form currents \(J(x)= J^a_\mu(x) T^a \dd x^\mu\) and \(\widetilde{J}(x)= \widetilde{J}^a_\mu(x) T^a \dd x^\mu\) as given by (\ref{current01}) and (\ref{current02}), and these satisfy the following conditions:}
\begin{itemize}
	\item[{\bf(i)}] The nonlocal currents are conserved in the perturbed theory, that is 
	\[
	\partial^\mu J_\mu^a = 0\; .
	\]
	\item[{\bf(ii)}] The nonlocal currents \({\cal L}^a(z, \bar{z})\) and \(\bar{{\cal R}}^a(z,\bar{z})\) become local for the undeformed theory, and they generate the Kac-Moody algebra of the WZNW theory. Equivalently, in the asymptotic limit \(\eta\to 0\) the currents  \({\cal L}^a_\ast(z)\) and \(\bar{{\cal R}}^a_\ast(\bar{z})\) satisfy the OPEs (\ref{ope1}) and (\ref{ope2}).
	\item[{\bf(iii)}] The components \({\cal L}^a(\tau, \sigma)\), \({\cal H}^a(\tau, \sigma)\), \(\bar{{\cal R}}^a(\tau, \sigma)\) and \({\cal K}^a(\tau, \sigma)\) do not generally give the massive current algebra as defined in \cite{Bernard:1990jw}.
\end{itemize}
\end{itembox}

Before continuing, we need to justify these hypotheses. Conditions {\bf(i)} and {\bf(ii)} are obvious. We are just saying that we have a conservation law associated to the nonlocal currents \(J\) and \(\tilde{J}\) and that in the undeformed limit we recover the WZNW model with Kac-Moody symmetry \(\hat{\mathfrak{g}}\times\hat{\mathfrak{g}}\). In fact, when we considered the condition \({\cal A}\simeq {\cal A}_\ast + {\cal O}(\eta)\) before, we secretly used the hypothesis {\bf(ii)}. Condition {\bf(ii)} also implies that in the conformal theory limit, the components \({\cal L}\) and \(\bar{\cal R}\) become local. Equivalently, we have assumed that once we deform the theory, one introduces these disorder operators, which in essence means that the currents become nonlocal. In the examples of Yang-Baxter deformations \cite{Kawaguchi:2010jg, Kawaguchi:2011pf, Kawaguchi:2011mz, Kawaguchi:2012ug, Kawaguchi:2013gma, Kawaguchi:2013lba} and in the cases studied by Bernard and LeClair \cite{Bernard:1990ys}, we can find specific cases of nonlocality generated by deformations (although in different contexts) which justify this second hypothesis.

Finally, condition {\bf(iii)} is subtler. We relax the conditions imposed by Bernard in \cite{Bernard:1990jw}, and that particularly says that we do not assume that the OPEs close solely in left- or right-moving modules of \(\mathfrak{g}\times \bar{\mathfrak{g}}\). In other words, we assume that any product involving the components \({\cal L}^a(t,x)\), \({\cal H}^a(t,x)\), \(\bar{{\cal R}}^a(t,x)\) and \({\cal K}^a(t,x)\) needs to be expanded in terms of these currents and their descendants. Evidently, the conditions considered in \cite{Luscher:1977uq, Bernard:1990jw, Hauer:1997ig, Hauer:1997gw} 
\be 
{\cal L} \bar{\cal R}\sim 0\; ,\quad {\cal L} {\cal K}\sim 0\; ,\quad {\cal H} \bar{\cal R}\sim 0 \; ,\quad {\cal H} {\cal K}\sim 0\; ,
\ee
are extremely useful, but it seems artificial in a very generic situation such as those we try to address. It is reasonable to assume that once the theory is deformed, the chiral and antichiral structure of the theory is spoiled, and it is this what condition {\bf(iii)} denotes. Additionally, Zamolodchikov \cite{Zamolodchikov:1989zs} imposes that the Hilbert space decomposition is preserved once we deform the theory, but it is not obviously generic enough, so we explore other possibilities.

One can now try to use these general hypotheses to study the algebra generated by nonlocal symmetries. It is evidently a highly nontrivial task, but we can start with a babe problem, 
finding some algebraic constraints imposed by physical arguments. That is exactly what we start doing now.

\section{Algebra of nonlocal charges}\label{algebra}

As we have just seen, the chiral decomposition of the theory is broken in the deformed theory, therefore we do not want to preserve the notation associated to structure as an artifact. In order to study the algebra generated by the nonlocal charges \(\mathcal{Q}^{1a}\) and \(\mathcal{Q}^{2\bar{a}}\), it is much more convenient to define the column vector
\be 
[\mathscr{J}^{\alpha}] := 
\begin{pmatrix}
J^a \\ 
\tilde{J}^{\bar{a}}
\end{pmatrix} = 
\begin{pmatrix}
J^a_+ \\ J^a_- \\ \tilde{J}^{\bar{a}}_+ \\ \tilde{J}^{\bar{a}}_-
\end{pmatrix} \equiv 
\begin{pmatrix}
{\cal L}^a \\ {\cal H}^a \\ {\cal K}^{\bar{a}} \\ \bar{\cal R}^{\bar{a}}
\end{pmatrix}
\; ,
\ee 
with \(\mathscr{J}^1 = J^a\) and \(\mathscr{J}^2 = \tilde{J}^{\bar{a}}\). Moreover, we already know that nonlocality is equivalently written as the equal-time braiding relations
\bse
\be 
\label{braidings}
\begin{split} 
[\mathscr{J}^{\alpha}(t,x_1)]\cdot [ \mathscr{J}^{\beta}(t,x_2)]= [\mathrm{R}^{\alpha \beta}(x_{12})]_{\gamma \delta} [\mathscr{J}^{\gamma}(t,x_2)]\cdot [ \mathscr{J}^{\delta}(t,x_1)]\; ,
\end{split}
\ee
where 
\be 
[\mathrm{R}^{\alpha \beta} ]\equiv 
\begin{pmatrix}
[\mathrm{R}^{\alpha \beta}]_{cd} & [\mathrm{R}^{\alpha \beta}]_{c\bar{d}}\\
[\mathrm{R}^{\alpha \beta}]_{\bar{c}d} & [\mathrm{R}^{\alpha \beta}]_{\bar{c}\bar{d}}
\end{pmatrix} \quad \Rightarrow  \quad 
[\mathrm{R}]\equiv 
\begin{pmatrix}
[\mathrm{R}^{ab}] & [\mathrm{R}^{a\bar{b}}]\\
[\mathrm{R}^{\bar{a}b}] & [\mathrm{R}^{\bar{a}\bar{b}}]
\end{pmatrix} 
\; .
\ee
\ese
Using the vertex operator representation, the braiding relations (\ref{braidings}) are closely related to anyonic statistics, which has been analyzed in many different contexts, including in the search for quantum group structures in two-dimensional conformal field theories. The relation between these ideas, if any, is not clear and deserves further investigation. 

\subsection{R-commutators}\label{r-commu}

Now we can start some full-fledged calculations. Remember that one of our main hypotheses is that these nonlocal objects mimic some features of the local currents. Associativity of their product imposes that the following expressions are equal
\bse
\begin{align}
\left( \mathscr{J}^{\alpha}(x_1) \mathscr{J}^{\beta}(x_2)\right) \mathscr{J}^{\gamma}(x_3) &= \mathrm{R}^{\alpha\beta}_{\delta \epsilon}(x_{12}) \mathrm{R}^{\epsilon\gamma}_{\zeta \kappa}(x_{13}) \mathrm{R}^{\delta \zeta}_{\eta \iota}(x_{23}) \mathscr{J}^{\eta}(x_3) \mathscr{J}^{\iota}(x_2) \mathscr{J}^{\kappa}(x_1)\\
\mathscr{J}^{\alpha}(x_1) \left(  \mathscr{J}^{\beta}(x_2)  \mathscr{J}^{\gamma}(x_3) \right) &=   \mathrm{R}^{\beta\gamma}_{\delta \epsilon}(x_{23}) \mathrm{R}^{\alpha \delta}_{\eta\zeta} (x_{13}) \mathrm{R}^{\zeta \epsilon}_{\iota \kappa}(\sigma_{12}) \mathscr{J}^{\eta}(x_3) \mathscr{J}^{\iota}(x_2) \mathscr{J}^{\kappa}(x_1) \; ,
\end{align}
\ese
and it immediately implies that the matrices \(\mathrm{R}^{ab}_{cd}\) are, as expected, solutions of the quantum Yang-Baxter equation
\be 
\mathrm{R}_{12}(x_{12})\mathrm{R}_{13}(x_{13}) \mathrm{R}_{23}(x_{23})  = \mathrm{R}_{23}(x_{23}) \mathrm{R}_{13}(x_{13})\mathrm{R}_{12}(x_{12})\; .
\ee
One may use the direct calculation -- that is, we take the explicit product of currents in the quantized theory -- to fix the matrix \(\mathrm{R}\), but currently it is obviously impossible to complete this program, and for now this matrix is an input of our construction. It might be possible to solve this conundrum by deepening our understanding of the consistency conditions, something along the lines of a Bootstrap program. Fortunately, when we have an integrable Yang-Baxter deformation, the situation is slightly better, and it seems possible to use the Yang-Baxter solution which defines the deformation as a seed for the braiding matrix. We discuss more of this possibility in the specific example of section (\ref{YangBaxter}).

The important aspect is that with the column currents \(\mathscr{J}^\alpha\), one can define the charges (\ref{charges}) as
\bse
\be 
\mathcal{Q}^{\alpha}  :=\frac{1}{2\pi i}\left( \int \dd z \mathscr{J}^{\alpha}_z + \int \dd \bar{z} \mathscr{J}_{\bar{z}}^{\alpha}  \right)\;,
\ee
and now we can invoke the results of Bernard and LeClair \cite{Bernard:1990ys} to compute what we call \emph{R-commutators}
\be 
\label{r-comm}
[\mathcal{Q}^{\alpha},\mathcal{Q}^{\beta}]_{\mathrm{R}} \equiv \mathcal{Q}^{\alpha}\mathcal{Q}^{\beta} - \mathrm{R}^{\alpha\beta}_{\gamma\delta} \mathcal{Q}^{\gamma} \mathcal{Q}^{\delta} = 
\widehat{\mathcal{Q}}^{\alpha}\left(\mathcal{Q}^{\beta} \right)\; ,
\ee
where the rhs of this equation is given by
\be 
\label{alg.charges}
\begin{split} 
\widehat{\mathcal{Q}}^{\alpha}\left(\mathcal{Q}^{\beta} \right) =  \frac{1}{(2\pi i)^2}& \left( \int \dd z \dd w \mathscr{J}_{z}^{\alpha}(z, \bar{z}) \mathscr{J}_w^{\beta}(w, \bar{w}) + \int \dd z \dd \bar{w} \mathscr{J}_z^{\alpha}(z, \bar{z}) \mathscr{J}_{\bar{w}}^{\beta}(w, \bar{w})\right.\\
& \left.+ \int \dd \bar{z} \dd w \mathscr{J}_{\bar{z}}^{\alpha}(z, \bar{z}) \mathscr{J}_w^{\beta}(w, \bar{w}) 
 + \int \dd \bar{z} \dd \bar{w} \mathscr{J}_{\bar{z}}^{\alpha}(z, \bar{z}) \mathscr{J}_{\bar{w}}^{\beta}(w, \bar{w})\right)\; .
\end{split}
\ee
In summary, the problem is then reduced to the calculation of four terms
\be 
\widehat{\cal Q}^{a}\left({\cal Q}^{\bar{b}} \right)\; , \quad 
\widehat{\cal Q}^{\bar{a}}\left({\cal Q}^{b} \right)\; , \quad 
\widehat{\cal Q}^{a}\left({\cal Q}^{b} \right)\; , \quad 
\widehat{\cal Q}^{\bar{a}}\left({\cal Q}^{\bar{b}} \right)\; .
\ee
\ese
Using the first element to understand the structure of these objects, we have
\be 
\begin{split}
\widehat{\cal Q}^{a}\left({\cal Q}^{\bar{b}} \right) = \frac{1}{2\pi i} & \left[ \int \dd z\ \underset{\scriptscriptstyle w\to z}{\mathrm{Res}}\left({\cal L}^a_\ast(z) {\cal K}^{\bar{b}}(w, \bar{w}) \right) + 
\int \dd \bar{z}\ \underset{\scriptscriptstyle \bar{w}\to \bar{z}}{\mathrm{Res}}\left( {\cal H}^a(z, \bar{z}) \bar{{\cal R}}^{\bar{b}}_\ast(\bar{w}) \right)\right. \\
& \left. +\frac{1}{2\pi i} \int \dd z \dd \bar{w} {\cal L}^a(z, \bar{z}) \bar{{\cal R}}^{\bar{b}}(w, \bar{w}) + \frac{1}{2\pi i} \int \dd \bar{z} \dd w {\cal H}^a(z, \bar{z}) {\cal K}^{\bar{b}}(w, \bar{w})  \right]\; .
\end{split}
\ee
where we have considered the perturbative expansion
\be 
{\cal L}(z, \bar{z})\simeq {\cal L}_\ast(z) + \eta {\cal L}_1(z, \bar{z})\; , \quad \bar{{\cal R}}(z, \bar{z})\simeq \bar{{\cal R}}_\ast(\bar{z}) + \eta \bar{{\cal R}}_1(z, \bar{z})\; .
\ee

The first constraint we need to impose is related to the obvious fact that in the undeformed limit \(\eta\to 0\) we must have \({\cal H}^a\to 0\) and \({\cal K}^{\bar{a}}\to0\). Consequently, one should suppose that these fields are of order \({\cal O}(\eta)\), which can be easily verified in (\ref{cons.laws03ha}) and (\ref{cons.laws03ka}). Additionally, we can rewrite the charges in terms of the coordinates \((x,t)\) instead of \((z, \bar{z})\); therefore
\bse
\be 
\begin{split}
\label{qq01}
\widehat{\cal Q}^{a}\left({\cal Q}^{\bar{b}} \right) =  \int \dd x \dd y \left( {\cal L}^a(x,t) {\cal K}^{\bar{b}}(y,t) \right.  & + 
{\cal H}^a(x, t) \bar{{\cal R}}^{\bar{b}}(y,t)  \\
 &\left. + {\cal L}^a(x, t) \bar{{\cal R}}^{\bar{b}}(y,t) + {\cal H}^a(x, t) {\cal K}^{\bar{b}}(y, t)  \right)\; ,
\end{split}
\ee
and we see that in order to finish this calculation we simply need to integrate the perturbed operators OPEs. Additionally, we also have 
\be 
\begin{split}
\widehat{\cal Q}^{a}\left({\cal Q}^{b} \right) =   \int \dd x \dd y  \left(  {\cal L}^a(x, t) {\cal L}^{b}(y, t) \right. & + {\cal L}^a(x, t) {\cal H}^{b}(y,t) \\ 
& \left.  + {\cal H}^a(x, t) {\cal L}^{b}(y, t) + {\cal H}^a(x, t) {\cal H}^{b}(y,t)  \right)\; ,
\end{split}
\ee
\be 
\begin{split}
	\widehat{\cal Q}^{\bar{a}}\left({\cal Q}^{b} \right) =  \int \dd x \dd y \left( {\cal L}^{\bar{a}}(x,t) {\cal K}^{b}(y,t) \right.  & + 
	{\cal H}^{\bar{a}}(x, t) \bar{{\cal R}}^{b}(y,t)  \\
	&\left. + {\cal L}^{\bar{a}}(x, t) \bar{{\cal R}}^{b}(y,t) + {\cal H}^{\bar{a}}(x, t) {\cal K}^{b}(y, t)  \right)\; ,
\end{split}
\ee
and evidently
\be 
\begin{split}
\widehat{\cal Q}^{\bar{a}}\left({\cal Q}^{\bar{b}} \right) =  \int \dd x \dd y  \left({\cal K}^{\bar{a}}(x, t) \bar{{\cal R}}^{\bar{b}}(y, t) \right. & + 
\bar{{\cal R}}^{\bar{a}}(x, t) {\cal K}^{\bar{b}}(y, t)\\ 
& \left.+ \bar{{\cal R}}^{\bar{a}}(x, t) \bar{{\cal R}}^{\bar{b}}(y, t) + {\cal K}^{\bar{a}}(x, t) {\cal K}^{\bar{b}}(y, t)  \right)\; .
\end{split}
\ee
\ese

We have just seen that the algebra of nonlocal charges can be completely determined from the analysis of operator product expansions of the corresponding conserved currents. In other words, we need to find the coefficients of the matrix
\be 
[\mathscr{J}^\alpha\mathscr{J}^{\beta}] =
[\mathscr{J}^{\alpha}]\otimes [\mathscr{J}^{\beta}] = 
\begin{pmatrix}
\violet{{\cal L}^a {\cal L}^b} & \purple{{\cal L}^a {\cal H}^b} & \purple{{\cal L}^a{\cal K}^{\bar{b}}} & \purple{{\cal L}^a \bar{\cal R}^{\bar{b}}}\\
\purple{{\cal H}^a {\cal L}^b} & \magenta{{\cal H}^a {\cal H}^b} & \magenta{{\cal H}^a {\cal K}^{\bar{b}}} & \purple{{\cal H}^a \bar{\cal R}^{\bar{b}}}\\
\purple{{\cal K}^{\bar{a}} {\cal L}^b} & \magenta{{\cal K}^{\bar{a}} {\cal H}^b} & \magenta{{\cal K}^{\bar{a}} {\cal K}^{\bar{b}}} & \purple{{\cal K}^a \bar{\cal R}^{\bar{b}}}\\
\purple{\bar{\cal R}^{\bar{a}} {\cal L}^b} & \purple{\bar{\cal R}^{\bar{a}} {\cal H}^b} & \purple{\bar{\cal R}^{\bar{a}} {\cal K}^{\bar{b}}} & \violet{\bar{\cal R}^{\bar{a}} \bar{\cal R}^{\bar{b}}}
\end{pmatrix} \; ,
\ee
where consistency imposes that the lowest orders in \(\eta\) are
\be
\label{ope.exp}
\begin{split} 
\violet{{\cal L}{\cal L}}\; , \violet{\bar{\cal R}\bar{\cal R}} & \sim {\cal O}(\eta^0)\\
\purple{{\cal L}{\cal H}}\; , \purple{{\cal H}{\cal L}} & \; ,
\purple{{\cal L}{\cal K}} \; , \purple{{\cal K}{\cal L}}\; ,
\purple{{\cal L}\bar{\cal R}}\; , \purple{\bar{\cal R}{\cal L}}\; ,
\purple{{\cal H}\bar{\cal R}}\; , \purple{\bar{\cal R}{\cal H}}\; ,
\purple{{\cal K}\bar{\cal R}}\; ,\purple{\bar{\cal R} {\cal K}}  \sim {\cal O}(\eta)\\
\magenta{{\cal H}{\cal H}}\; , \magenta{{\cal H}\bar{\cal K}} & \; , 
\magenta{{\cal K}\bar{\cal H}}\; , \magenta{{\cal K}\bar{\cal K}} \sim {\cal O}(\eta^2)\; .
\end{split}
\ee
Observe that there is another subtlety at this point. The matrix above cannot be reduced to an upper (or lower) triangular matrix since we do not have the usual bosonic or fermionic statistics at our disposal anymore; therefore we need to compute all OPEs. In \cite{Bernard:1990ys} the authors considered that the OPE closes independently in the left of right modules so that terms of the form \({\cal L}\bar{\cal R}\) are regular. As we said before, that simplification is very unnatural, and we consider the most general case. The next section is an attempt to unveil properties of these products.

\subsection{(Nonlocal) Operator Product Expansion}

Before starting our calculations, we need to remark -- maybe redundantly -- that the present construction is intrinsically quantum; and in this sense, all these expressions are inside correlation functions. Therefore, it should be clear that once we deform the theory, a new prescription for the calculations of vacuum expectation values (VEVs) must be imposed, and it is well known that the perturbed VEVs are related to those of the original theory by
\be 
\langle \Phi(z_1, \cdots, z_n) \rangle = \langle \Phi(z_1, \cdots, z_n) e^{-\eta \int {\cal O}_{pert.}} \rangle_0\; .
\ee
where \(\Phi(z_1, \cdots, z_n)\) is an arbitrary product of fields and \(\langle \cdots \rangle_0\) is the undeformed VEV. In this sense, we simply want to develop methods to calculate the correlators when they have insertions of nonlocal operators.

We have seen that once we deform the theory, nonlocal conserved charges and consequently their corresponding currents can be defined. For integrable theories, the existence of classical nonlocal charges is often associated to the underlying Yangian symmetry. Much more important, there exist prescriptions to build an infinite tower of nonlocal charges \({\cal Q}_{(n)}\), \(n\geq 1\), starting from a local charge \({\cal Q}_{(0)}\), see \cite{Bernard:1992ya, MacKay:2004tc, Loebbert:2016cdm}. The algebra of these charges can be calculated using the Poisson bracket (which can be later quantized) and it has been shown that the computation of the brackets of \({\cal Q}_{(0)}\) and \({\cal Q}_{(1)}\) (together with the so-called Serre relations) are enough to determine the algebraic structure of this symmetry. 

What we meant is that local and nonlocal currents often come together, but in the present work we study a more exotic situation where we do not specifically consider any (obvious) local charge to be paired with the nonlocal ones, in such a way that the more familiar Yangian symmetries are generated. So, we need to learn how to live with these nonlocal charges, and in particular we need to describe their algebra directly from the OPEs, perhaps following ideas similar to those used in the vertex operator algebras construction. But how can we do that? The most direct approach is to find the classical counterpart of the nonlocal charges defined in the previous section, take the Poisson bracket, and finally use the standard canonical quantization. This direct program is not only difficult, it is unpractical and unrealistic. Alternatively, given that the construction so far has been intrinsically quantum (we considered marginally relevant operators, OPEs and so on), we need to take it as an advantage to make \emph{Ans\"atze} for the OPEs, and try to see what the physical constraints have to say about the coefficients.

As we said many times before, nonlocality is comprised in the existence of an \(\mathrm{R}\)-matrix, which imposes that for each pair of operators \({\cal O}_1(x)\) and \({\cal O}_2(0)\), one says that they are mutually nonlocal if under permutation their product \({\cal O}_1(x_1){\cal O}_2(0)\) transforms as  \(\exp(2\pi i \alpha){\cal O}_2(0){\cal O}_1(x_1)\) where the phase \(\exp(2\pi i \alpha)\) is an unspecified component of the \(\mathrm{R}\)-matrix. Observe that this situation is similar to what happens with anyons or parafermions \cite{Mussardo:2010mgq, Fateev:1985mm} statistics, where the anyonic or parafermionic pair of operators would be \(\alpha\)-nonlocal if under an analytic continuation of \(x\), their product \({\cal O}_1(x_1){\cal O}_2(0)\) gets a phase \(\exp(2\pi i \alpha)\). Even more importantly, if we embrace this analogy for the case of \(\mathbb{Z}_N\) invariant lattice models, we have the order \(\sigma_k\) and disorder \(\mu_k\) operators, with \(k=1, \dots, N-1\) which are mutually nonlocal. For the definition of the OPEs one needs to introduce the parafermionic operators themselves, and consequently we do not have any ambiguity to express the operator product expansions.

\begin{figure}[ht]
	\begin{center}
		\includegraphics[scale=0.40]{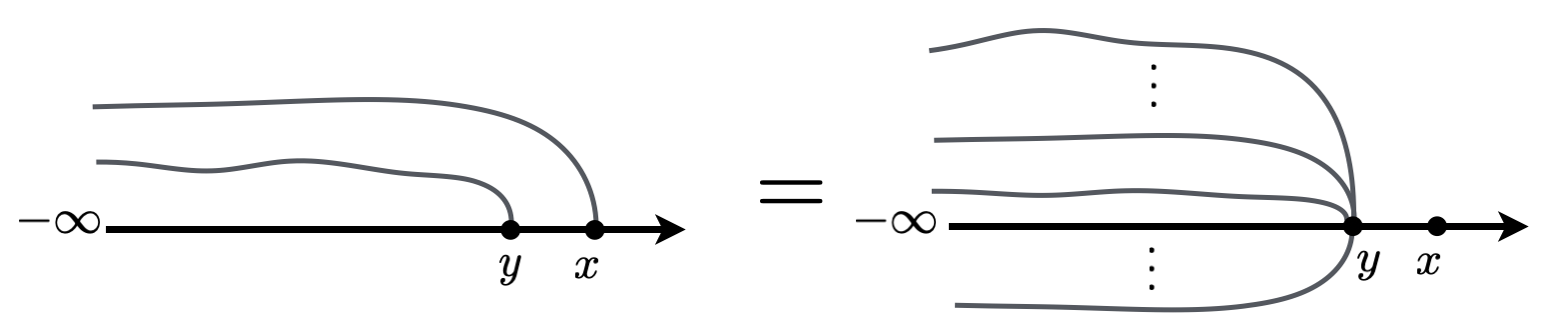}	
		\caption{Nonlocal OPEs in a simply connected space.}
		\label{nlope}
	\end{center}
\end{figure}

The nonlocal nature of these operators makes it hard to give an interpretation in terms of operator product expansion, but we assume that when the space is simply connected, two nonlocal operators which are close to each other can be replaced by a sum of nonlocal operators. Observe that in a complete general situation where we know the deformed local and nonlocal operators, we could impose that the OPEs would close using both types of operators, there is no contradiction to assume that. The fact is that we have just nonlocal operators to care about, so let us assume that these operators are enough to make a closed algebra.

\begin{figure}[hb]
\begin{center}
\includegraphics[scale=0.40]{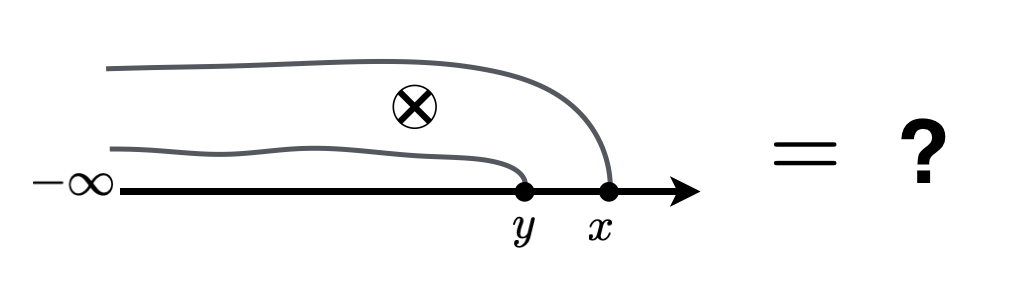}	
\caption{Nonlocal OPEs in a non simply connected space.}
\label{nlope2}
\end{center}
\end{figure}

 Moreover, we have also assumed that the space is simply connected. That is very important because we need to assume that the integration curves are all homotopic, see figure \ref{nlope}. In fact, for homotopic curves we can use the expansion of figure \ref{nlope}, but it is not exactly clear how we can define the OPEs for two operators defined via homotopically inequivalent curves; see figure \ref{nlope2}. We do not address this type of subtlety in this paper, so let us assume that we have a simply connected space where the first construction is appropriate.

All in all, the OPEs are defined in the usual way, except we drop the consistency conditions implied by locality \cite{Hauer:1997gw, Hauer:1997ig}. They have the following structure,
\bse 
\be 
\mathscr{J}^{\alpha}(z,\bar{z}) \mathscr{J}^{\beta}(w,\bar{w})\sim \sum_{\gamma k} \mathscr{C}^{\alpha \beta}_{k \gamma}(|z-w|) \mathscr{O}_k^{\gamma}(w, \bar{w})\; ,
\ee
where \(k\) is an enumeration index. Given that we have assumed that the only Lie algebra-valued operators available are the nonlocal currents, we have that \(\mathscr{O}_k^{\gamma}\) are the currents themselves \(\mathscr{J}^{\gamma}\) and derivatives thereof. Hence, the relevant terms are
\be 
\mathscr{J}^{\alpha}(z,\bar{z}) \mathscr{J}^{\beta}(w,\bar{w})\sim {\cal C}^{\alpha \beta}_{\gamma}(|z-w|) \mathscr{J}^{\gamma}(w,\bar{w}) + 
{\cal D}^{\alpha \beta}_{\gamma}(|z-w|)(\partial \mathscr{J})^{\gamma}(w,\bar{w})\; .
\ee
Moreover, it is convenient to use the coordinates \((t,x)\); then
\be 
\mathscr{J}_m^{\alpha}(x) \mathscr{J}_n^{\beta}(0)\sim ({\cal C}_{mn }^p)^{\alpha \beta}_{\gamma}(x) \mathscr{J}_p^{\gamma}(0) + 
({\cal D}_{mn}^{pq})^{\alpha \beta}_{\gamma}(x)\partial_p \mathscr{J}_q^{\gamma}(0)\; .
\ee
As before, the index \(\alpha=(a, \bar{a})\) denotes the left- and right-moving Lie algebras and now we make a small change of notation in the indices \((m,n)\), namely
\be 
J^a \equiv  (J_+^a, J_-^a) = ({\cal L}^a, {\cal H}^a)\; \quad 
\tilde{J}^{\bar{a}} \equiv  (\tilde{J}_+^{\bar{a}}, \tilde{J}_-^{\bar{a}}) = ({\cal K}^{\bar{a}}, \bar{{\cal R}}^{\bar{a}})\; ,
\ee
and \(\mathscr{J}\) remains the same \(\mathscr{J}^{\alpha} \equiv \left(J^a, \tilde{J}^{\bar{a}}\right) \). As we see, this small modification enhances the readability of the results. Additionally, we further assume the expansion
\be 
\boxed{
\label{ope}
\mathscr{J}_m^{\alpha}(x) \mathscr{J}_n^{\beta}(0)\sim \mathscr{F}^{\alpha \beta}_{\gamma}  \left( {\cal C}^p_{mn}(x) \mathscr{J}_p^{\gamma}(0) + {\cal D}_{mn}^{pq}(x)\partial_p \mathscr{J}_q^{\gamma}(0) \right)} \; .
\ee
\ese

Initially one can impose some obvious constraints related to the engineering dimensions, perturbative expansion in \(\eta\) and associativity of the product. More specifically, we have the following:

\paragraph{Engineering dimensions:} The easiest constraint comes from dimensional analysis. As in \cite{Hauer:1997gw}, we have 
\be 
 {\cal C}^p_{mn}(x) \simeq {\cal O}(|x|^{-1-0})\; , \quad
  {\cal D}^{pq}_{mn}(x) \simeq {\cal O}(|x|^{-0})\; ,
\ee
where \(|x|^{-0}\) denotes possible logarithmic divergences. 

\paragraph{The structure constants:} The first interesting constraints are related to the structure constants \(\mathscr{F}^{\alpha \beta}_{\gamma}\), \( {\cal C}^p_{mn} \) and \({\cal D}^{pq}_{mn}\). The OPEs \(\mathscr{J}^{a} \mathscr{J}^{b}\) and \(\mathscr{J}^{\bar{a}} \mathscr{J}^{\bar{b}}\) give the Kac-Moody algebra for \(\eta=0\); therefore we consider expansions of the form 
\be 
\begin{array}{lll}
\mathscr{F}^{ab}_{c} \simeq f^{ab}_{\phantom{ab}c} + \eta  \tilde{f}^{ab}_{\phantom{ab}c} &\quad &  \mathscr{F}^{\bar{a}\bar{b}}_{\bar{c}} \simeq f^{\bar{a}\bar{b}}_{\phantom{\bar{a}\bar{b}}\bar{c}} + \eta  \tilde{f}^{\bar{a}\bar{b}}_{\phantom{\bar{a}\bar{b}}\bar{c}} \\
\mathscr{F}^{ab}_{\bar{c}} \simeq \eta  \tilde{f}^{ab}_{\phantom{ab}\bar{c}} &\quad &  \mathscr{F}^{\bar{a}\bar{b}}_{c} \simeq \eta  \tilde{f}^{\bar{a}\bar{b}}_{\phantom{\bar{a}\bar{b}}c}\\
\mathscr{F}^{\bar{a}b}_{\gamma} \simeq \eta  \tilde{f}^{\bar{a}b}_{\phantom{\bar{a}b}\gamma} &\quad &  \mathscr{F}^{a\bar{b}}_{\gamma} \simeq \eta  \tilde{f}^{a\bar{b}}_{\phantom{\bar{a}b}\gamma}\qquad \gamma=c, \bar{c}\; .
\end{array}
\ee
Using the OPEs structure (\ref{ope.exp}) we have that the following coefficients 
\bse
\be
\begin{split} 
&{\cal C}^+_{+-}\; , {\cal C}^+_{-+}\; ,  {\cal C}^-_{--}\; ,  {\cal C}^+_{-+}\; ,  {\cal C}^-_{-+}\; , {\cal C}^+_{+-} \; , {\cal C}^-_{+-} \; , {\cal C}^-_{-+}\\
&{\cal D}^{-+}_{+-}\; , {\cal D}^{++}_{+-}\; , {\cal D}^{-+}_{-+}\; , {\cal D}^{++}_{-+} \; , {\cal D}^{--}_{--} \; , {\cal D}^{+-}_{--} \; , {\cal D}^{--}_{-+} \; , {\cal D}^{-+}_{-+} \; , {\cal D}^{+-}_{-+} \; , {\cal D}^{++}_{-+}\; ,\\ 
& {\cal D}^{--}_{++}\; ,  {\cal D}^{+-}_{++} \; ,  {\cal D}^{--}_{+-} \; ,  {\cal D}^{+-}_{+-}
 \; ,  {\cal D}^{--}_{-+}  \; ,  {\cal D}^{+-}_{-+}
\end{split}
\ee
are of order \({\cal O}(\eta)\). Additionally, the coefficients 
\be
\begin{split} 
&{\cal C}^+_{--}\;  , {\cal C}^-_{++}  \\
&{\cal D}^{-+}_{--}\; , {\cal D}^{++}_{--}\; ,  {\cal D}^{+-}_{++} \; ,  {\cal D}^{--}_{++}
\end{split}
\ee
are of order \({\cal O}(\eta^2)\) and they can be neglected in the current perturbative analysis. All other coefficients are the leading order \({\cal O}(\eta^0)\) terms.
\ese

Moreover, let us denote the inverse of \((\mathscr{F}_\gamma)^{\alpha\beta}\) by \((\mathscr{F}^\gamma)_{\alpha\beta}\), that is
\be 
(\mathscr{F}_\gamma)^{\alpha\beta}(\mathscr{F}^\gamma)_{\alpha'\beta}=\delta^\alpha_{\alpha'}\; ,
\ee 
and using the inverse structure constants, and the expression (\ref{ope}) we obtain
\be 
\boxed{
\mathscr{F}^\gamma_{\alpha\beta}\mathscr{J}_m^{\alpha}(x) \mathscr{J}_n^{\beta}(0)\sim {\cal C}^p_{mn}(x) \mathscr{J}_p^{\gamma}(0) + 
{\cal D}_{mn}^{pq}(x)\partial_p \mathscr{J}_q^{\gamma}(0) } \; .
\ee

\paragraph{Associativity:} Associativity imposes that the braiding matrix satisfies the Yang-Baxter equation, but if we use this condition in the OPEs, we find an additional important constraint. Specifically we need to study the two sides of the equality
\be 
\left( \mathscr{J}_m^{\alpha}(x) \mathscr{J}_n^{\beta}(y)\right) \mathscr{J}_r^{\lambda}(z) = 
\mathscr{J}_m^{\alpha}(x)\left( \mathscr{J}_n^{\beta}(y) \mathscr{J}_r^{\lambda}(z)\right)\; ,
\ee
where we (unwisely) use \(z\) as a real parameter, and it must not be confused with the complex coordinate we used previously. The left-hand side of this equation gives
\bse 
\be 
\begin{split}
\left( \mathscr{J}_m^{\alpha}(x) \mathscr{J}_n^{\beta}(y)\right) & \mathscr{J}_r^{\lambda}(z)  = \\ & = \mathscr{F}^{\alpha \beta}_{\gamma} \mathscr{F}^{\gamma \lambda}_{\sigma} \left\{ 
\left[ 
 {\cal C}^p_{mn}(x-y) {\cal C}^s_{pr}(y-z) +  {\cal D}^{pq}_{mn}(x-y) \partial^y_p {\cal C}^s_{qr}(y-z)
\right]\mathscr{J}_s^{\sigma}(z)
\right.  \\
& +\left. \left[
{\cal C}^p_{mn}(x-y) {\cal D}^{st}_{pr}(y-z) +  {\cal D}^{pq}_{mn}(x-y) \partial^y_p {\cal D}^{st}_{qr}(y-z)
\right]\partial_s \mathscr{J}_t^{\sigma}(z)\right\}\; ,
\end{split}
\ee
and the right-hand side gives
\be 
\begin{split}
\mathscr{J}_m^{\alpha}& (x)  \left( \mathscr{J}_n^{\beta}(y) \mathscr{J}_r^{\lambda}(z) \right)  = \\ & =  \mathscr{F}^{\beta\lambda}_{\gamma} \mathscr{F}^{\alpha\gamma}_{\sigma} \left\{ 
\left[ 
 {\cal C}^p_{nr}(y-z) {\cal C}^s_{mp}(x-z) +  {\cal D}^{pq}_{nr}(y-z) \partial^y_p {\cal C}^s_{mq}(x-z)
\right]\mathscr{J}_s^{\sigma}(z)
\right.  \\
+ & \left[
{\cal C}^p_{nr}(y-z) {\cal D}^{st}_{mp}(x-z) +  {\cal D}^{sq}_{nr}(y-z) {\cal C}^{t}_{mq}(x-z) + 
{\cal D}^{pq}_{nr}(y-z) \partial^y_p {\cal D}^{st}_{mq}(x-z)\right]\partial_s \mathscr{J}_t^{\sigma}(z)\\
+ &\left. 
{\cal D}^{pq}_{nr}(y-z) {\cal D}^{st}_{mq}(x-z)\partial^z_p \partial_s^z  \mathscr{J}_t^{\sigma}(z)
\right\}\; .
\end{split}
\ee
\ese
Comparing both sides we have that the term with two derivatives \(\partial^2\mathscr{J}\) must be paired with regular terms in the expansion. Additionally we have
\bse 
\be
\boxed{
\begin{aligned}
& {\cal C}^p_{mn}(x-y) {\cal C}^s_{pr}(y-z) +  {\cal D}^{pq}_{mn}(x-y) \partial^y_p {\cal C}^s_{qr}(y-z)= \\ 
& \hspace{1.5cm } = {\cal C}^p_{nr}(y-z) {\cal C}^s_{mp}(x-z) +  {\cal D}^{pq}_{nr}(y-z) \partial^z_p {\cal C}^s_{mq}(x-z)
\end{aligned}}
\ee
and 
\be 
\boxed{
\begin{aligned}
& {\cal C}^p_{nr}(y-z) {\cal D}^{st}_{mp}(x-z) +  {\cal D}^{sq}_{nr}(y-z) {\cal C}^{t}_{mq}(x-z) + 
{\cal D}^{pq}_{nr}(y-z) \partial^y_p {\cal D}^{st}_{mq}(x-z) = \\
&\hspace{1.5cm} 
 = {\cal C}^p_{nr}(y-z) {\cal D}^{st}_{mp}(x-z) +  {\cal D}^{sq}_{nr}(y-z) {\cal C}^{t}_{mq}(x-z) + 
{\cal D}^{pq}_{nr}(y-z) \partial^y_p {\cal D}^{st}_{mq}(x-z)\; .
\end{aligned}}
\ee
\ese

\paragraph{Crossing symmetry:} Finally, we should also impose crossing symmetry. We do not open the calculations in full details, but we impose the following conditions,
\be 
\contraction{}{\mathscr{J}_m^{\alpha}}{}{ \mathscr{J}_n^{\beta}}
\contraction{\mathscr{J}_m^{\alpha} \mathscr{J}_n^{\beta}}{\mathscr{J}_p^{\gamma}}{}{ \mathscr{J}_q^{\delta}}
\mathscr{J}_m^{\alpha} \mathscr{J}_n^{\beta}\mathscr{J}_p^{\gamma} \mathscr{J}_q^{\delta} 
\sim 
\contraction{}{\mathscr{J}_m^{\alpha}}{\mathscr{J}_n^{\beta}}{\mathscr{J}_p^{\gamma}}
\contraction[2ex]{\mathscr{J}_m^{\alpha}}{\mathscr{J}_n^{\beta}}{\mathscr{J}_p^{\gamma}}{ \mathscr{J}_q^{\delta}}
\mathscr{J}_m^{\alpha} \mathscr{J}_n^{\beta}\mathscr{J}_p^{\gamma} \mathscr{J}_q^{\delta}
\sim 
\contraction[2ex]{}{\mathscr{J}_m^{\alpha}}{\mathscr{J}_n^{\beta}\mathscr{J}_p^{\gamma}}{\mathscr{J}_q^{\delta}}
\contraction{\mathscr{J}_m^{\alpha}}{\mathscr{J}_n^{\beta}}{}{\mathscr{J}_p^{\gamma}}
\mathscr{J}_m^{\alpha} \mathscr{J}_n^{\beta}\mathscr{J}_p^{\gamma} \mathscr{J}_q^{\delta}
\ee
where the symbols \(\sim\) denote the fact we need to take into account the braiding relations.

These two previous conditions, namely the associativity and crossing symmetries, are extremely important but difficult to be properly analyzed. 

\subsection{Discrete symmetries}

As in \cite{Hauer:1997gw}, we could try to consider the constraints imposed by discrete symmetries but, as we said before, the nonlocality of these currents is closely related to the anyonic statistics which has as one of its essential features the violation of parity and time reversal transformations. Given that the current understanding of these objects is not complete, it is not clear to the author if we should impose these constraints, although CPT and C symmetries are unlikely to be violated.

In this respect, we see that the introduction of proportionality constants under these transformations is the safest route for us in the present paper. Moreover, a possible CPT violation of the currents should not be confused with inconsistencies of the theory. These nonlocal operators are not related to any observables in the final theory, and we should think of them as computational devices. We may remember that ghost fields are also nonphysical, since they violate the spin statistics, but they are also useful computational devices. Modulo this subtlety, these constraints mean the following relations

\paragraph{\bf \S 1.} Parity \(\mathsf{P}\) is defined by
\bse
\begin{align}
\mathsf{P} \mathscr{J}_m^{\alpha}(t, x) \mathsf{P}^{-1} & \overset{?}{=} (-)^m {\cal P}_m^n \mathscr{J}_n^{\alpha}(t,- x) \\
\mathsf{P} \partial_m \mathscr{J}_n^{\alpha}(t, x) \mathsf{P}^{-1} & \overset{?}{=} (-)^{m+n} {\cal P}_{m n}^{r s}\partial_r \mathscr{J}_s^{\alpha}(t,- x)\; .
\end{align}
\ese
for \(m,n,r,s=0,1\). Observe that in the usual case we have \({\cal P}_m^n=\delta^n_m\) and \({\cal P}_{m n}^{r s}=\delta^r_m \delta^s_n\), but we have allowed nontrivial terms to denote our ignorance on the nonlocal terms.

\paragraph{\bf \S 2.} Similarly, the \(\mathsf{CPT}\) symmetry gives
\bse
\begin{align}
\Theta \mathscr{J}_m^{\alpha}(t, x) \Theta ^{-1} & \overset{?}{=} - \theta_m^n \mathscr{J}_n^{\alpha}(-t,- x) \\
\Theta \partial_m \mathscr{J}_n^{\alpha}(t, x) \Theta^{-1} & \overset{?}{=} \theta_{m n}^{r s} \partial_r \mathscr{J}_s^{\alpha}(-t,- x)
\end{align}
\ese
where \(\theta_m^n=\delta^n_m\) and \(\theta_{m n}^{r s}=\delta^r_m \delta^s_n\) in the case of local charges. These relations imply
\be 
\boxed{ \theta_p^v \mathcal{C}_{mn}^p(x) = -   \theta_m^r  \theta_n^s\mathcal{C}_{rs}^p(-x) \qquad  \theta^{uv}_{pq}\mathcal{D}_{mn}^{pq}(x) = \theta_m^r  \theta_n^s \mathcal{D}_{rs}^{pq}(-x)}\; .
\ee

\paragraph{\bf \S 3.} The charge conjugation \(\mathsf{C}\) seems to be an unquestionable constraint, and it goes as follows:
\be 
\mathsf{C} \mathscr{J}_m^{\alpha}(t, x) \mathsf{C}^{-1}= B^\alpha_{\phantom{\alpha}\beta}\mathscr{J}_m^{\beta}(t, x)\; ,
\ee
where one should also impose
\bse
\be 
\mathsf{C}^2 \mathscr{J}_m^{\alpha} (\mathsf{C}^{-1})^2= \mathscr{J}_m^{\alpha} 
\ \ \Rightarrow \ \ \boxed{  B^\alpha_{\phantom{\alpha}\beta} B^\beta_{\phantom{\alpha}\gamma} = \delta^\alpha_{\phantom{\alpha} \gamma}}
\ee
and
\be
\begin{split}
 & \mathscr{F}^\gamma_{\alpha\beta}\mathsf{C}\mathscr{J}_m^{\alpha} \mathscr{J}_n^{\beta}\mathsf{C} ^{-1} =\\ 
 & \qquad = \mathsf{C} \left( 
 \mathcal{C}^p_{mn}(x)\mathscr{J}^\gamma_p(0) + 
 \mathcal{D}^{pq}_{mn}(x) \partial_p \mathscr{J}^\gamma_q(0)
 \right)\mathsf{C}^{-1}
  \ \ \Rightarrow \ \ 
\boxed{  \mathscr{F}^\gamma_{\alpha\beta} B^\alpha_{\phantom{\alpha}\lambda} B^\beta_{\phantom{\alpha}\eta} = \mathscr{F}^\zeta_{\lambda\eta} B^\gamma_{\phantom{\gamma}\zeta} }\; .
\end{split}
\ee
\ese

In order to be clear, we may think of this situation where we have several operators associated to discontinuities, such as a bunch of Dirac strings, and that these operators are conserved and satisfy an algebra. When inserted into correlation functions these operators impose conditions on the observables, but are not by themselves observables of the theory. We may see that depending on the properties \({\cal P}\) and \(\theta\) many interesting situations may happen. For example, triviality conditions appear when there is an odd number of currents inside a correlation function, and the matrices \({\cal P}\) and \(\theta\) are different from the identity. In any case, if these ideas are physically possible or if they are wild speculation will be the theme of a future publication, for now we want to see other possible constraints.

\subsection{Braiding relations}

Finally, we now turn our attention to constraints imposed by nonlocality itself. As we said many times before, the braiding matrix \(\mathrm{R}\) comprises the nonlocality of the currents, and consequently the constraints on the algebra of charges. Then
\be 
\mathscr{J}_m^{\alpha}(x) \mathscr{J}_n^{\beta}(0) = R^{\alpha\beta}_{\gamma\delta}(x) \mathscr{J}_n^{\gamma}(0) \mathscr{J}_m^{\delta}(x) \; ,
\ee
and one can easily show that
\bse
\be 
\label{cond.01}
{\cal C}_{mn}^p(x) = \left(- {\cal F}^\lambda _{\alpha\beta} R^{\alpha\beta}_{\gamma\delta} {\cal F}_\lambda ^{\gamma\delta}\right) {\cal C}_{nm}^p(x)\; .
\ee
Observe that this condition imposes that the diagonal terms vanish unless 
\be 
{\cal F}^\lambda _{\alpha\beta} R^{\alpha\beta}_{\gamma\delta} {\cal F}_\lambda ^{\gamma\delta}=-1\; ,
\ee 
and we can justify this choice by imposing that in the CFT limit the Kac-Moody algebra is recovered. The same condition also imposes that
\be 
\label{cond.02}
\mathcal{D}_{mn}^{pq}(x) =  x^p  {\cal C}_{nm}^q(x) - \mathcal{D}_{nm}^{pq}(x)\; .
\ee
\ese

Let us now define the tensor
\be 
\mathcal{T}_{mn}^{pq}(x):= \mathcal{D}_{mn}^{pq}(x) - \frac{1}{2}x^p  {\cal C}_{mn}^q(x)\; ,
\ee
and from it we can readily show that (\ref{cond.02}) can be equivalently written as
\be 
\boxed{{\cal T}^{pq}_{mn}(x) = -  {\cal T}_{nm}^{pq}(x)}\; ,
\ee
where we evidently used the symmetry of \({\cal C}_{mn}^q(x)\). We have just settled the antisymmetry of the indices \((m,n)\). Now, we can be utterly pragmatical and write \(\mathcal{T}_{mn}^{pq}(x)\) in terms of its symmetric, skewsymmetric and diagonal in the indices \((p,q)\), that is,
\be 
{\cal T}^{pq}_{mn}(x) = {\cal T}^{(pq)}_{mn}(x) + {\cal T}^{[pq]}_{mn}(x) + \eta^{pq}{\cal T}_{mn}(x)\; .
\ee
The first and obvious observation is that current conservation imposes that the diagonal terms in \((p,q)\) are identically \(0\), that is \({\cal T}_{mn}=0\). All in all, we have the expansion
\be 
\begin{split}
\mathscr{F}^\gamma_{\alpha\beta}\mathscr{J}_m^{\alpha}(x) \mathscr{J}_n^{\beta}(0)\sim {\cal C}^p_{mn}(x) & \mathscr{J}_p^{\gamma}(0)\\ 
& + \left( {\cal T}^{(pq)}_{mn}(x) + {\cal T}^{[pq]}_{mn}(x)  +\frac{1}{2} x^p{\cal C}^q_{mn}(x)  \right)\partial_p \mathscr{J}_q^{\gamma}(0) \; .
\end{split}
\ee
Quite remarkably, similar conditions can be found using the currents locality \cite{Luscher:1977uq, Bernard:1990ys, Hauer:1997ig}. As a direct consequence of these calculations, we can import the early results of \cite{Luscher:1977uq, Bernard:1990ys, Hauer:1997ig} to determine the OPE coefficients of the nonlocal OPEs. Therefore, these terms can be written as
\bse 
\begin{align}
\mathcal{C}_{mn}^p(x) & = C_1 \eta_{mn} x^2 x^p + C_2 (x_m \delta_n^p + x_n\delta_m^p) + C_3 x_m x_n x^p\\
T_{mn}^{(pq)} & = t_s \left( x^p (x_m \delta_n^q - x_n\delta_m^q) +  x^q (x_m \delta_n^p - x_n\delta_m^p) \right)\\
T_{mn}^{[pq]} & = t_a \left( \delta_m^p \delta_n^q - \delta_m^q \delta_n^p \right) x^2\; ,
\end{align}
\ese
where one can use the current conservation
\be 
\boxed{\partial^m \mathcal{C}^p_{mn} = 0 \qquad \partial^m \mathcal{D}^{pq}_{mn} = 0}\; .
\ee
to determine a series of algebraic equation for the coefficients \(C_i\), \(t_s\) and \(t_a\). We refer to \cite{Hauer:1997gw} for the explicit expression of those functions. 

\subsection{Algebra of charges}

In an ideal situation where all the constraints above are completely solved, we would have all necessary ingredients to finish our calculations for the algebra of charges we started in section \ref{r-commu}. For example, we know that 
\be 
\mathcal{Q}^a \mathcal{Q}^{\bar{b}}- \mathrm{R}^{a \bar b}_{\bar c d} \mathcal{Q}^{\bar c} \mathcal{Q}^{d} = \widehat{\mathcal{Q}}^a( \mathcal{Q}^{\bar{b}})\; ,
\ee
where we need to calculate the terms
\bse
\be 
\begin{split}
\widehat{\mathcal{Q}}^a( \mathcal{Q}^{\bar{b}}) & =  \sum_{\substack{m,n,p,q\\= +,-}}  
\eta \tilde{f}_{\phantom{a\bar b}c}^{a\bar b}\int \dd x \dd y \left( 
\mathcal{C}^p_{mn}(|x-y|)  \mathscr{J}_p^{c}(y) + \mathcal{D}^{pq}_{mn}(|x-y|) \partial_q  \mathscr{J}_q^{c}(y) \right) + \\
& + \sum_{\substack{m,n,p,q\\= +,-}}  \eta
\tilde{f}_{\phantom{a\bar b}\bar c}^{a\bar b}\int \dd x \dd y \left(
\mathcal{C}^p_{mn}(|x-y|)  \mathscr{J}_p^{\bar c}(y) + \mathcal{D}^{pq}_{mn}(|x-y|) \partial_q  \mathscr{J}_q^{\bar c}(y)\; .
\right)
\end{split}
\ee
Moreover, \(\widehat{\mathcal{Q}}^{\bar{a}}( \mathcal{Q}^b)\) is obtained from a simple transformation \(a\to \bar{a}\) and \(\bar{b}\to b\) in the equation above. Additionally
\be 
\begin{split}
\widehat{\mathcal{Q}}^a( \mathcal{Q}^{b}) & =  \sum_{\substack{m,n,p,q\\= +,-}}  (f_{\phantom{a b} c}^{ab}+\eta\tilde{f}_{\phantom{a b}c}^{ab})
\int \dd x \dd y \left( 
\mathcal{C}^p_{mn}(|x-y|)  \mathscr{J}_p^{c}(y) + \mathcal{D}^{pq}_{mn}(|x-y|) \partial_q  \mathscr{J}_q^{c}(y) \right) + \\
& + \sum_{\substack{m,n,p,q\\= +,-}}  
\eta \tilde{f}_{\phantom{ab}\bar c}^{ab}\int \dd x \dd y \left(
\mathcal{C}^p_{mn}(|x-y|)  \mathscr{J}_p^{\bar c}(y) + \mathcal{D}^{pq}_{mn}(|x-y|) \partial_q  \mathscr{J}_q^{\bar c}(y)\; ,
\right)
\end{split}
\ee
and finally
\be 
\begin{split}
\widehat{\mathcal{Q}}^{\bar{a}}( \mathcal{Q}^{\bar{b}}) & =  \sum_{\substack{m,n,p,q\\= +,-}}  \eta\tilde{f}_{\phantom{\bar{a} \bar{b}}c}^{\bar a\bar b}\int \dd x \dd y \left( 
\mathcal{C}^p_{mn}(|x-y|)  \mathscr{J}_p^{c}(y) + \mathcal{D}^{pq}_{mn}(|x-y|) \partial_q  \mathscr{J}_q^{c}(y) \right) + \\
& + \sum_{\substack{m,n,p,q\\= +,-}}  
(f_{\phantom{\bar a\bar b}\bar c}^{\bar a\bar b} + \eta \tilde{f}_{\phantom{\bar a\bar b}\bar c}^{\bar a\bar b})\int \dd x \dd y \left(
\mathcal{C}^p_{mn}(|x-y|)  \mathscr{J}_p^{\bar c}(y) + \mathcal{D}^{pq}_{mn}(|x-y|) \partial_q  \mathscr{J}_q^{\bar c}(y)
\right)\; .
\end{split}
\ee
\ese
The final \(\mathrm{R}\)-commutators are
\be 
\label{algeb}
\boxed{\mathcal{Q}^\alpha \mathcal{Q}^{\beta}- \mathrm{R}^{\alpha \beta}_{\gamma \delta} \mathcal{Q}^{\gamma} \mathcal{Q}^\delta = \mathscr{F}^{\alpha\beta}_{\lambda} \mathcal{T}^\lambda\; , }
\ee
where 
\be 
\boxed{\mathcal{T}^\lambda = 
\int \dd x \dd y \left(
\mathcal{C}^p_{mn}(|x-y|)  \mathscr{J}_p^{\lambda}(y) + \mathcal{D}^{pq}_{mn}(|x-y|) \partial_q  \mathscr{J}_q^{\lambda}(y)
\right)\; .}
\ee

Unfortunately we do not have any means to finish this calculation, but the problem has been posed. In summary, if we want to calculate the algebra of charges for a given theory with nonlocal operators, we just need to understand two specific problems. The first problem is to elucidate the behavior of the braiding \(\mathrm{R}\)-matrix jiggling around all these calculations. At this point it is just an input and, as we see in the next section, there is a good candidate for these objects in the case of Yang-Baxter deformations.

The second problem is to determine the behavior of objects \(\mathcal{T}^\lambda\). One immediate question to be answered is to explain when and if \(\mathcal{T}^\lambda\) can be written in terms of the charges \(\mathcal{Q}^\alpha\). Additionally, one may also try to see if there is any other interesting possibility. For example, remember that in a quantum theory it is known that if an operator \(\mathcal{Q}\) satisfies the following relation
\be 
[\mathcal{T}^\lambda, \mathcal{Q}^\beta]=\mathcal{T}^\lambda[\mathcal{Q}]\mathcal{Q}^\alpha\; ,
\ee
we say that \(\mathcal{T}^\lambda[\mathcal{Q}]\) is the topological charge of the operator \(\mathcal{Q}\). Therefore, it would be most interesting to verify if and when these objects are topological-like charges.

\section[Applications: From Yang-Baxter to TT \& TJ deformations]{Applications: From Yang-Baxter to \(T\bar T\) \&  \(T\bar J\) deformations} \label{string}

We finish our paper with a rather incomplete analysis of how the central objects of the current work can be applied when we have deformations of the worldsheet and boundary CFTs. It is evidently an unfortunate situation that we cannot describe the whole mechanism in full generality, but we can try to see some features of this construction. Although the topic of the present paper is not necessarily confined in the string theory context, these two examples are inside this research program.

Before starting our analysis, let us quote an important result that classifies the types of deformations we are interested in. Chaudhuri and Schwartz \cite{Chaudhuri:1988qb} and recently Borsato and Wulff \cite{Borsato:2018spz} have considered marginal deformations of the form
\be 
\label{mar.def}
{\cal O}(z, \bar{z}) = c_{a\bar{b}}{\cal L}^a(z) \bar{\cal R}^{\bar{b}}(\bar{z})  \; ,
\ee
where \(c_{a\bar{b}}\) are constant coefficients. They have shown that in exactly marginal deformations the constants \(c_{a\bar{b}}\) identify two maximally solvable subalgebras. See also \cite{Orlando:2006cc} for a systematic collection of results and examples. Knowing this result we try to see what it can do for us. 

\subsection{Yang-Baxter deformations}\label{YangBaxter}

Consider integrable deformations of AdS\(_3\times \mathbb{S}^3\). The first thing one may observe is that being an exactly marginal deformation is generally not enough to guarantee that the deformed background is a string solution. Quite generally, there are several other conditions, amongst them the existence of a nilpotent BRST (Becchi-Rouet-Stora-Tyutin) operator in the deformed theory which may be used to define physical states of the quantized strings. 

Observe that when we say string solution we strictly mean that the theory has as massless excitations only those fields of the usual ten-dimensional supergravities. In this sense, marginal deformations that give string backgrounds must be built from physical vertex operators of the original theory. Considering the Yang-Baxter deformations \cite{Delduc:2013qra, Kawaguchi:2014qwa, Arutyunov:2015mqj, Wulff:2016tju}, consistency of the vertex operator is translated into a condition on the classical r-matrix called unimodularity \cite{Borsato:2016ose}, and this same condition has also been discussed in the pure spinor formalism in \cite{Bedoya:2010qz, Mikhailov:2012id}.

Therefore, given a Yang-Baxter deformation one has the following possibilities:
\begin{itemize}
	\item {\bf Unimodular deformations:} As we have just said, these deformations give consistent string theory solutions; therefore, they are obtained from exactly marginal operators and on top of that, the unimodularity condition is satisfied. That also means that such deformations satisfy the so-called (weak) Chaudhuri-Schwartz (CS) conditions defined in \cite{Chaudhuri:1988qb}, see also \cite{Orlando:2006cc, Araujo:2018rho, Borsato:2018spz}.
	\item {\bf Nonunimodular deformations:} In this case we do not find consistent string theory solutions, but two interesting things may happen:
		\begin{enumerate}
			\item These deformations may be generated by marginally relevant operators. In this case, the current-current deformation does not satisfy the CS condition. This condition is equivalent to deformations generated by classical r-matrices that do not satisfy the unimodularity conditions, and the symmetry generators do not belong to the maximal soluble Lie algebra \cite{Borsato:2018spz}. The nonunimodularity is realized in terms of a new vector field \(K\) in the massless spectrum of the theory, and this term adds to the beta functions a \(K^2\) contribution.
			\item But it may also happen that deformations are exactly marginal (the CS condition is satisfied, which means that the deformed CFT has vanishing beta functions) but that there is a massless nonphysical field in the string spectra \cite{Bedoya:2010qz, Mikhailov:2012id, Araujo:2017enj} which manifests itself again as the  Killing vector field \(K\) at the supergravity level. That essentially means that the unimodularity condition is not satisfied, despite the fact the deformation is exactly marginal. Observe that this case is precisely what happens in \cite{Wulff:2018aku}, where the  solutions have null Killing vector fields, i.e. \(K\) with \(K_\mu K^\mu=0\). Given that the modifications of the beta functions are realized as a term of the form \(K^2\), the beta functions are still vanishing, and consequently the resulting theory is still a CFT, but with a nonphysical field \(K\) that spoils their string theory interpretation. That essentially means that other stringy requirements are not satisfied.
		\end{enumerate}
\end{itemize}

These are the types of deformations we study now.

\subsubsection*{Marginally relevant deformations of the AdS\(_3\) strings}

Among the extensive collection of advantages of this particular deformation, one of them is extremely compelling to our proposal; namely, there is a very natural candidate for the braiding matrix \(\mathrm{R}\). In other words, suppose we have Yang-Baxter deformations of nonlinear sigma models based on the \(\mathrm{R}\)-matrix \(\mathrm{R}\simeq 1 + \eta r +\mathscr{O}(\eta^2)\), where \(r\) satisfies the classical Yang-Baxter equation. Therefore, one can use this same \(\mathrm{R}\)-matrix as a seed for our braiding \(\mathrm{R}\)-matrix. More specifically, for a deformation of the form
\be 
\delta S\propto \int r_{a\bar{b}}{\cal L}^a(z) \bar{\cal R}^{\bar b}(\bar{z})\; ,
\ee
with \(r_{a\bar{b}}=-r_{\bar{b}a}\), one can impose that \(\mathrm{R}_{\alpha\beta}\simeq (1 + \eta r)_{\alpha\beta}\). Observe that in the infinitesimal limit we consider now, the diagonal components (in the left- and right-moving sectors \(\mathrm{R}_{ab}\) and \(\mathrm{R}_{\bar{a}\bar{b}}\)) do not play any role in the deformation itself, but are important in the algebra of nonlocal charges.

One may try to study these nonlocal objects for deformations of string theory in the background AdS\(_3\times \mathbb{S}^3\times T^4\) supported with a B-field. In order to make the text self-contained, we review some of its features in the Appendix \ref{ads3}. It is well known that this theory is equivalent to an  \(SL(2, \mathbb{R})\times SU(2)\)-WZNW model, and this particular model has two maximally solvable subalgebras generated by \(\{V_3, V_-, K_3\}\) and \(\{\bar{V}_3, \bar{V}_-, \bar{K}_3\}\). Therefore, any perturbation defined by these generators is exactly marginal. 

We have a list of possible marginally relevant deformations of the \(SL(2, \mathbb{R})\) WZNW model
\be
\begin{split}
\delta_{-+} S &= \frac{\eta}{2\pi} \int \dd^2z {\cal L}^-(z) \bar{\cal R}^+(\bar{z})\; , \quad \delta_{+-} S= \frac{\eta}{2\pi} \int \dd^2z {\cal L}^+(z) \bar{\cal R}^-(\bar{z})\\
\delta_{3+} S &= \frac{\eta}{2\pi} \int \dd^2z {\cal L}^3(z) \bar{\cal R}^+(\bar{z})\; , \quad \delta_{+3} S= \frac{\eta}{2\pi} \int \dd^2z {\cal L}^+(z) \bar{\cal R}^3(\bar{z})\\
\delta_{++} S &= \frac{\eta}{2\pi} \int \dd^2\sigma {\cal L}^+(z) \bar{\cal R}^+(\bar{z})\; ,
\end{split}
\ee
where the currents are
\be
\begin{split}
{\cal L}^3&= k \left( \partial u + \cosh 2 \rho \partial v \right)\; , \qquad {\cal L}^\pm = k \left(\partial \rho \pm i \sinh 2 \rho \partial v \right) e^{\mp i 2 u}\; , \\
{\cal R}^3&= k \left( \bar\partial v + \cosh 2 \rho \bar\partial u \right)\; , \qquad {\cal R}^\pm = k \left(\bar\partial \rho \pm i \sinh 2 \rho \partial u \right) e^{\mp i 2 v}\; ,
\end{split}
\ee
and the familiar OPEs are given by
\begin{align} 
{\cal L}^3(z) {\cal L}^\pm(w)& \sim \pm \frac{{\cal L}^\pm(w)}{z-w} & & {\cal R}^3(\bar z) {\cal R}^\pm(\bar w) \sim \pm \frac{{\cal R}^\pm(\bar w)}{\bar z-\bar w}\nn \\
{\cal L}^3(z) {\cal L}^3(w) & \sim - \frac{k}{2(z-w)^2} &&  {\cal R}^3(\bar z) {\cal R}^3(\bar w) \sim - \frac{k}{2(\bar z-\bar w)^2} \\
{\cal L}^+(z) {\cal L}^-(w) & \sim \frac{k}{(z-w)^2} + \frac{2 {\cal L}^3(w)}{z-w} && 
{\cal R}^+(\bar z) {\cal R}^-(\bar w) \sim \frac{k}{(\bar z-\bar w)^2} + \frac{2 {\cal R}^3(\bar w)}{\bar z-\bar w} \nn\; .
\end{align}

It has been argued in \cite{Borsato:2018spz} that the coefficients \(c^{ab}\) may be interpreted as the nondiagonal elements of  an r-matrix one can engineer as a solution of the classical Yang-Baxter equation. In particular, the example {\bf 4.3.4} of \cite{Sakamoto:2018krs} corresponds to the r-matrix
\be 
\label{pert.}
r=-\frac{1}{\sqrt{2}}V_+\wedge \left( V_3 + \bar{V}_3\right) \; \quad \Leftrightarrow \quad \; \delta_{+3} S= \frac{\eta}{2\pi} \int \dd^2z {\cal L}^+(z) \bar{\cal R}^3(\bar{z})\; ,
\ee
and we notice that the perturbation by the operator \({\cal O}(z,\bar{z})=\eta {\cal L}^+(z) \bar{\cal R}^3(\bar{z}) \) is not only marginally relevant, but also (classically) integrable by the arguments of \cite{Borsato:2018spz}. With this deformation we can now (try to) apply the techniques we developed in previous sections regarding the existence and algebraic relations of the nonlocal charges.

As we said before, the deformed nonlocal fields are defined in terms of a new gauge-valued field \({\cal A}\) obtained from its dynamical equations (\ref{dyn.eq.a}) or, equivalently, from (\ref{dyn.eq.b}). For example, if we assume the marginally relevant operator \({\cal O}(z,\bar{z})=\eta {\cal L}^+(z) \bar{\cal R}^3(\bar{z}) \), it is straightforward to show that equations (\ref{dyn.eq.b}) are
\bse 
\be 
\partial \bar{\partial} X^+  = \eta X^+ Y^+ \; , \quad
\partial \bar{\partial} X^-  = - \eta X^+ Y^-\; , \quad
\partial \bar{\partial} X^3 = 0
\ee
and
\be 
\partial \bar{\partial} Y^+  = 0 \; , \quad
\partial \bar{\partial} Y^-  = 2 \eta X^3 Y^3\; , \quad
\partial \bar{\partial} Y^3 = -\eta X^+ Y^3\; ,
\ee
where \(X^a\equiv \partial {\cal A}^a\) and  \(Y^a\equiv \bar{\partial} {\cal A}^a\) for \(a=\pm, 3\). With the solution for these equations for the components \({\cal A}^a\) we can directly characterize the nonlocal currents \({\cal L}^a\) and \(\bar{{\cal R}}^a\) and their algebraic properties.
\ese

\subsection[Comments on holographic TT and TJ ] {Comments on holographic \(T \bar T\) and \(T \bar J\)}

Certain irrelevant and integrability preserving deformations, called \(T\bar{T}\) and \(T\bar{J}\) deformations, have been studied initially in \cite{Zamolodchikov:2004ce, Caselle:2013dra, Smirnov:2016lqw, Cavaglia:2016oda, Guica:2017lia}. As we know, the subject of the present work is essentially marginal deformations of field theories; consequently, the perturbations defined by Zamolodchikov, Smirnov and Guica may seem absolutely unrelated to the current discussion. Here the magic of AdS/CFT comes into play. It has been suggested that certain marginal deformations of the string worldsheet theory, which is defined by the gravity dual to the undeformed holographic CFT, encompass all essential features of these \(T\bar{T}\) and \(T\bar{J}\) deformations \cite{Giveon:2017nie, Giveon:2017myj, Chakraborty:2018vja, Apolo:2018qpq, Chakraborty:2019mdf, Giveon:2019fgr}.

More specifically, given a 2D QFT, not necessarily conformal, we can define the following irrelevant operator of weight \((2,2)\),
\be 
T\bar{T}(x)=\lim_{y\to x}\left(T(y)\bar{T}(x) - \Theta(y)\bar{\Theta}(x)  \right)\; ,
\ee
where \(T\), \(\bar{T}\), \(\Theta\) and \(\bar{\Theta}\) are the components of the stress-energy tensor, which must satisfy the conservation law
\be
\partial_{\bar{x}}T=\partial_x \bar{\Theta}\; , \quad \partial_x \bar{T} = \partial_{\bar{x}}\Theta.
\ee 
Observe that this irrelevant operator is universal in the sense it can always be defined, since this object only depends on the stress-energy tensor.

When we have a theory with (at least) a \(SL(2, \mathbb{R})\times U(1)\) symmetry, one can define the following \((2,1)\)-irrelevant operator
\be 
J\bar{T}(x)=\lim_{y\to x}\left(J(y)\bar{T}(x) - \bar{J}(y)\Theta(x)  \right)\; ,
\ee
where the conservation law for the current must also be satisfied 
\be 
\partial \bar{J} +\bar{\partial} J=0\; .
\ee

From the holographic viewpoint, we may also study this deformation from a stringy perspective. As usual, when the two-dimensional CFT is holographic dual to a string background of the form AdS\(_3\times \mathcal{N}\), where \({\cal N}\) is a compact space,  its \(T\bar{T}\) and \(T\bar{J}\) deformations above should correspond to a double trace perturbation. Consequently, these deformations change the boundary conditions of the theory \cite{McGough:2016lol, Bzowski:2018pcy}. Quite remarkably, the authors of \cite{Giveon:2017nie, Giveon:2017myj, Chakraborty:2018vja, Apolo:2018qpq}, see also \cite{Giveon:2019fgr}, argue that there are single trace deformations of the AdS\(_3\) strings which reproduce the essential (or all?) features of the \(T\bar{T}\) and \(T\bar{J}\) deformations. These single trace operators are exactly what we need to implement our ideas on the existence and algebra of the nonlocal charges.

Let us concentrate on the \(T\bar{T}\) deformations since the \(T\bar{J}\) follows similar reasoning. It has been shown that the  \(T\bar{T}\) perturbation in the boundary CFT is equivalent, from the worldsheet perspective, to
\be 
\delta \mathscr{L}_{ws} = -\eta {\cal L}^-(z)\bar{\cal R}^-(\bar{z})\; .
\ee
One should notice that this transformation is, indeed, exactly marginal since it does not belong to the maximally soluble subalgebra specified in the previous section. Evidently, in this particular case, the limitation of our analysis is much more noticeable. 

In the Wakimoto representation of the AdS\(_3\) strings, we have the Lagrangian
\be 
\mathscr{L}_{ws}  = \beta\bar{\partial} \gamma + \bar{\beta}\partial \bar{\gamma} + \partial \phi \bar{\partial}\phi -\sqrt{\frac{2}{k}}\hat{R}\phi - e^{-\sqrt{\frac{2}{k}}\phi}\beta\bar{\beta}\; ,
\ee
where we have written explicitly the \(\beta\gamma\)-system which may be appropriately bosonized as
\bse
\be 
\begin{split}
& \gamma  = i \phi_- \qquad  \beta = i \partial \phi_+ \\
& \bar{\gamma} = i \bar{\phi}_- \qquad  \bar{\beta} = i \bar{\partial} \bar{\phi}_+ \quad \textrm{where} \quad \phi_+(z)\phi_-(w) \simeq \ln(z-w)\; ,
\end{split}
\ee
\ese
and now we can use the Coulomb gas formulation machinery. One can think of the bosonized fields in terms of the scalar field \(\Phi(z) = \phi_+(z) + \phi_-(z)\). 

The boundary \(T\bar{T}\) deformed theory is equivalent to the worldsheet theory
\be 
\mathscr{L}_{ws}  = \beta\bar{\partial} \gamma + \bar{\beta}\partial \bar{\gamma} + \partial \phi \bar{\partial}\phi -\frac{2}{\alpha_+}\hat{R}\phi -\left(\eta + e^{-\frac{2}{\alpha_+}\phi}\right)\beta\bar{\beta}\; ,
\ee
 in such a way that this new theory is not \(SL(2, \mathbb{R})\times SL(2, \mathbb{R})\) invariant anymore, therefore one could expect that the nonlocality of the deformed theory should be summarized in terms of the new scalar field \(\tilde{\Phi} = \Phi +\lambda (\texttt{nonlocal})\). Unfortunately, we have not been able to define the expressions for the nonlocal terms for exactly marginal deformations. The recent discussions of these issues (at the classical level) in terms of \(O(d,d)\) transformations may be the correct direction to address this problem, and it is an interesting direction to explore \cite{Araujo:2018rho, Orlando:2019rjg}.
 
At any rate, one needs to ensure that the net effect of the nonlocality is summarized in the algebraic discussion of section \ref{algebra}. The fact that the resulting theory is still a CFT imposes several constraints to the braiding matrix and in particular, we have the following OPE product matrix
\be 
[\mathscr{J}^\alpha\mathscr{J}^{\beta}] =
[\mathscr{J}^{\alpha}]\otimes [\mathscr{J}^{\beta}] = 
\begin{pmatrix}
\violet{{\cal L}^a {\cal L}^b} & 0 & 0 & 0\\
0 & 0 & 0 & 0\\
0 & 0 & 0 & 0\\
0 & 0 & 0 & \violet{\bar{\cal R}^{\bar{a}} \bar{\cal R}^{\bar{b}}}
\end{pmatrix} \; ,
\ee
since \({\cal H}^a= {\cal K}^{\bar{a}}=0\), and evidently we have the usual decomposition between holomorphic and antiholomorphic sectors in the Hilbert space. As a direct consequence we have the regularity of the following OPEs \({\cal L}^a \bar{\cal R}^{\bar{b}}=0\) and \(\bar{\cal R}^{\bar{a}} {\cal L}^b=0\). These constraints greatly simplify the computation of the \(\mathrm{R}\)-commutators (\ref{r-comm})
\be 
\begin{split}
[{\cal Q}^a, {\cal Q}^b ]_{\mathrm{R}} & = \widehat{\cal Q}^a({\cal Q}^b)\\
[{\cal Q}^a, {\cal Q}^{\bar{b}} ]_{\mathrm{R}} & = [{\cal Q}^{\bar{a}}, {\cal Q}^{b} ]_{\mathrm{R}} = 0\\
[{\cal Q}^{\bar{a}}, {\cal Q}^{\bar{b}} ]_{\mathrm{R}} & = \widehat{\cal Q}^{\bar{a}}({\cal Q}^{\bar{b}})\; .
\end{split}
\ee

As we said before, the methods we tried to define in this paper do not seem to be strong enough to address exactly marginal deformations. As a sad consequence, we are unable to compute, or at least to give a way to calculate the nonlocal charges above. Incurring the risk of repeating ourselves, the \(J\bar{J}\) deformations are better described by \(O(d,d)\) transformations, and recently the deformed theory nonlocality has been addressed in \cite{Orlando:2019rjg}; we see that integrability also plays an important role there. We can certainly explore those directions to understand the quantum algebra of these deformations, and we hope to study these points in a future work. 

\section{Conclusions}

In spite of their ubiquitousness in field theories, nonlocal charges are still difficult to explore, especially due to the absence of general methods to define them. In integrable theories, these are usually related to the Yangian symmetry which is believed to be an alternative to the inverse scattering method in the study of integrability. The role played by nonlocal charges, and the circumstances under which we can define them are not clear for generic quantum theories. In this paper we tried to address these problems using very unsophisticated and primitive methods, such as OPEs and brute force constraints imposed by physical arguments. Perhaps with future development of string theory and nonperturbative methods in field theories, one may use its methods to address these questions properly. 

After a brief review in section 2, we started our analysis of nonlocal charges in section 3 with the definition of currents directly from the deformations of free scalar representations of conformal theories. This method is a simple generalization of what has been done by LeClair and Bernard \cite{Bernard:1990ys}. On the other hand, the applicability of this approach is very limited and it represents an evident drawback of this technique. 

In the same section we dared to apply the idea to the currents themselves, and not in the fields defining them. These objects are written in terms of a new algebra valued field \({\cal A}\), and from it we addressed the algebraic properties of the deformed charges in section 4. We tried to find a handful of constraints imposed by physical conditions. Most of the analysis we performed in this work is extremely incipient, and should be obviously extended or modified to be useful in the precise definition of the algebra of nonlocal charges. 

The case of marginally relevant operators is easier to study, in particular because we were able to use some known methods developed by Zamolodchikov in the 80s \cite{Zamolodchikov:1989zs}, but even in this simplified context the calculations become involved very fast. In section 5 we tried to show how the nonlocal charges may appear in the calculation of pure NS-NS AdS\(_3\) integrable deformations in type-IIB string theory.In order to complete the calculations we need to solve a system of PDE for the coefficients \({\cal A}^a\), \(a=\pm,3\), which are important in the definition of the deformed currents. It is just a technical problem that may eventually be solved.

When the deformation is exactly marginal the situation is even harder. In this particular scenario, the deformed objects \({\cal L}\) and \(\bar{\cal R}\) become nonlocal under deformation as expected, but they preserve their chirality, and it necessarily implies that \({\cal H}^a(t,x)= 0\) and \({\cal K}^{\bar{a}}(t,x)=0\) despite the fact that \(\eta\neq 0\). 

Another important point we did not discuss in the current paper is the role of these charges, their physical meaning and how they act on the theory itself. As a matter of fact, these problems are poorly defined, but one may use as a guide the recent ideas on the role of Yangians in the action and correlation functions of the \({\cal N}=4\) super Yang-Mills \cite{Beisert:2018zxs, Beisert:2018ijg, Loebbert:2018lxk}. At any rate, it is not obvious to what extent those methods can be generalized to other theories.

There are many doable and immediate problems to be addressed in a finite time. Most of the calculations of section 3 should be properly done, in particular for the cases of nonsimply laced Lie algebras. This is, evidently, a rather mechanical work but in the process of opening the calculations one may have interesting insights. Much more important is to look for a stronger prescription for the definition of the \(\mathrm{R}\)-matrix. In the present work we considered it as an input, but it is evidently a very unsatisfactory situation, since it should be a consequence of the deformation itself, as in our example of Yang-Baxter deformations. Another interesting aspect is to understand if we can consider the right-hand side of equation (\ref{algeb}) as topological charges, if we can write it in terms of the original nonlocal charges, or even a more exotic situation.

For integrable deformation, there are better tools to handle the existence and properties of the nonlocal charges \cite{Orlando:2019rjg}, but for generic deformations one may try to use (and extend) some of the ideas we explored in this paper. New ideas are certainly needed, and we hope to report new results in future publications. 

\paragraph{Acknowledgements}

I am supported by the Swiss National Science Foundation under Grant No. \textsc{PP00P2\_183718/1}. This project has also been partly supported by the Korea Ministry of Education, Science and  Technology, Gyeongsangbuk-Do and Pohang City Independent Junior Research Groups at the Asia Pacific Center for Theoretical Physics. I also have been fortunate to have as collaborators Kentaroh Yoshida, Eoin \'O Colg\'ain, Thiago Fleury, Shahin Sheik Jabbari and Hossein Yavartanoo; without their expertise on related projects, the present work would certainly not exist. I am grateful to them for many discussions regarding several topics present in the current manuscript. I also thank Roberto Tateo for discussions.

\appendix 

\section{AdS\(_3\) Strings}\label{ads3}

In order to make the paper self-contained, we review the main features of strings propagating in the AdS\(_3\times S^3\) background supported by a B-field. We start with the sigma-model description of this geometry\footnote{Comparing with the notation of \cite{Borsato:2018spz}, we have \(u\equiv x^+\) and \(v\equiv x^-\). }
\bse 
\begin{align}
\dd s^2 & = \frac{1}{z^2}\left( \dd z^2 - \dd u \dd v \right) + \frac{1}{4}\left[ \dd \phi_3^2 + \cos^2 \phi_3 \dd \phi_1^2 +\left( \dd \phi_2 +\sin\phi_3\dd \phi_1 \right)^2 \right]\\
B & = \frac{\dd u\wedge \dd v}{z^2} - \frac{1}{2}\sin\phi_3 \dd \phi_1 \wedge \dd \phi_2\; .
\end{align}
\ese
Isometries of the AdS\(_3\) metric are defined by the following Killing vectors
\bse
\be 
\begin{split}
	k_0 & =u \partial_u+ \frac{z}{2}\partial_z \; , \quad k_{+1}=\partial_u \; , \quad k_{-1}= - u^2 \partial_u - z^2 \partial_v - u z \partial_z \\
	\tilde{k}_0 & =-v \partial_v- \frac{z}{2}\partial_z \; , \quad \tilde{k}_{-1}=\partial_v \; , \quad \tilde{k}_{+1}= - v^2 \partial_v - z^2 \partial_u - v z \partial_z \; ,
\end{split}
\ee
which satisfy the following commutation relations
\be 
[k_a, k_b]=-f_{ab}^{\phantom{ab}c}k_c\; , \quad [\tilde{k}_a, \tilde{k}_b]=-f_{ab}^{\phantom{ab}c}\tilde{k}_c
\ee
\ese
where \(f_{ab}^{\phantom{ab}c}\) are the structure constants of two copies of the algebra sl\((2, \mathbb{R})\), 
\be 
\begin{split}
	[s_0,s_{\pm 1}] & =\pm s_{\pm 1}\; , \quad [s_{+1}, s_{-1}] = 2s_0 \\
	[\tilde{s}_0, \tilde{s}_{\pm 1}] & =\pm \tilde{s}_{\pm 1}\; , \quad [\tilde{s}_{+1}, \tilde{s}_{-1}] = 2\tilde{ s}_0 \; .
\end{split}
\ee

For each Killing vector field above, we have the transformation (no sum in the index \(a\))
\bse
\be
\begin{split}
	\delta_a X^m & = \epsilon_a k_a(X^m)\\
	\delta_a G & = \epsilon_a  \mathscr{L}_{k_a}(G)=0 \\
	\delta_a B & = \epsilon_a \mathscr{L}_{k_a} B \\ 
	& = \epsilon_a \frac{1}{2}\left(k_a^p \partial_p B_{mn} +  \partial_m k_a^p B_{p n} +  \partial_n k_a^p B_{m p} \right) \dd X^m\wedge \dd X^n \equiv - \epsilon_a \dd J_a \; ,
\end{split}
\ee 
and 
\be 
\begin{split}
	\tilde{\delta}_a X^m & = \tilde{\epsilon}_a \tilde{k}_a(X^m)\\
	\tilde{\delta}_a G & = \tilde{\epsilon}_a  \mathscr{L}_{\tilde{k}_a}(G)=0 \\
	\tilde{\delta}_a B & = \tilde{\epsilon}_a \mathscr{L}_{\tilde{k}_a} B \\
	& = \tilde{\epsilon}_a\frac{1}{2}\left(\tilde{k}_a^p \partial_p B_{mn} +  \partial_m \tilde{k}_a^p B_{p n} +  \partial_n \tilde{k}_a^p B_{m p} \right) \dd X^m\wedge \dd X^n  \equiv - \tilde{\epsilon}_a \dd \tilde{J}_a\; ,
\end{split}
\ee
where \(\mathscr{L}_v\) is the Lie derivative along the vector \(v\).
\ese

The worldsheet action
\begin{subequations}
	\be 
	S=\frac{T}{2}\int \dd^2\sigma \partial X^m(G_{mn} + B_{mn})\bar{\partial} X^n\; 
	\ee
	explicitly gives
	\be 
	S=\frac{k}{2\pi}\int\dd^2 \sigma \left( 
	\frac{1}{z^2}\left(\partial z \bar{\partial} z - \partial v \bar \partial u \right) + 
	\frac{1}{4} \partial \phi_i \bar{\partial }\phi_i + \frac{1}{2}\partial \phi_2 \partial \phi_1 \sin\phi_3
	\right)\; ,
	\ee
	where we have used the following definitions for the worldsheet coordinates:
	\be 
	\begin{split}
		\sigma& =\sigma^0+\sigma^1\; , \; \bar{\sigma}=\sigma^0-\sigma^1 \; , \; \partial=\partial_\sigma \; , \;\bar{\partial}=\partial_{\bar{\sigma}} \\
		\eta^{\sigma\bar{\sigma}}& =\eta^{\bar{\sigma}\sigma}=-2\; , \; \epsilon^{\sigma\bar{\sigma}}=-\epsilon^{\bar{\sigma}\sigma}=-2\; ,\; \dd \sigma=\frac{1}{2}\dd \sigma \dd \bar{\sigma}\; .
	\end{split}
	\ee
\end{subequations}

The Noether currents are obtained from the variation
\be 
\label{var}
\delta_a S=\frac{T}{2}\int \dd^2\sigma \left[ ( \partial \epsilon_a) k_a^m(G_{mn} + B_{mn})\bar{\partial} X^n +  
\partial X^m (G_{mn} + B_{mn}) k_a^n (\bar{\partial} \epsilon_a) + \partial X^m \delta_a B_{mn} \bar{\partial} X^n\right]\; .
\ee
We can simplify the notation of this expression with the following currents
\be 
\begin{split} 
	\mathbb{J}_a\equiv \mathbb{J}_{a,+} & = k_a^m(G_{mn} - B_{mn})\partial X^n\\
	\bar{\mathbb{J}}_a\equiv \mathbb{J}_{a,-} & = k_a^m(G_{mn} + B_{mn})\bar{\partial} X^n\; .
\end{split}
\ee
We also need to calculate the last term in (\ref{var}). Given that \(\mathscr{L}_a B=-\dd J_a\), 
\bse 
\be 
\begin{split} 
	\delta_a B= \frac{1}{2}\delta_a B_{mn}\dd X^m \wedge \dd X^n & = -\epsilon^a \dd \left( J_{a,n}\dd X^n \right) \\
	& = - \epsilon_a \dd \left(J_a \dd \sigma + \bar{J}_a \dd \bar{\sigma}\right)\\
	& = - \frac{1}{2}\epsilon^a \left( \partial_m J_{a,n} - \partial_n J_{a,m} \right)\dd X^m\wedge \dd X^n\; .
\end{split}
\ee
Moreover, we notice that
\be 
\begin{split}
	\delta_a B & = 2 \delta_a B_{mn}\partial X^m \bar{\partial} X^n \left( \frac{1}{2}\dd \sigma \wedge \dd \bar{\sigma}\right) \\
	& = 2 \partial X^m \delta_a B_{mn} \bar{\partial} X^n \dd^2 \sigma \equiv 2 \delta_a B_{\sigma\bar{\sigma}} \dd^2 \sigma\; .
\end{split}
\ee
\ese
Integrating this expression we have 
\bse 
\be
\begin{split}
	\int \dd^2 \sigma \partial X^m  \delta_a B_{mn} \bar{\partial} X^n & = \frac{1}{2}\int \delta_a B =\frac{1}{2}\int \dd \epsilon_a \wedge J_a \\
	& = \int \dd^2 \sigma \left(\partial \epsilon_a \bar{j}_a - \bar{\partial} \epsilon_a j_a \right) \; ,
\end{split}
\ee
where \(J_a=j_a\dd \sigma + \bar{j}_a\dd \bar{\sigma}\)  with
\be 
j_a = J_{a,m}\partial X^m \;, \quad \bar{j}_a =J_{a,m} \bar{\partial} X^m\; .
\ee
\ese

Putting all these facts together, we have 
\be 
\delta_a S=\frac{T}{2}\int \dd^2\sigma  \left[ ( \partial \epsilon_a)\left( \bar{\mathbb{J}}_a + \bar{j}_a \right)+  
(\bar{\partial} \epsilon_a) \left(\mathbb{J}_a-  j_a \right) \right]\; .
\ee
Therefore, the Noether's currents can be written as
\be 
{\cal J}_{a, \pm} = \mathbb{J}_{a,\pm} \pm j_{a, \pm}\; .
\ee
Similarly, we have 
\be 
\tilde{{\cal J}}_{a, \pm} = \tilde{\mathbb{J}}_{a,\pm} \pm \tilde{j}_{a, \pm}\; .
\ee

Strings propagating in the AdS\(_3\) background can also be described as an \(SL(2, \mathbb{R})\) WZNW model. We show now that the sigma model currents we obtain in both descriptions are equal, and it settles the equivalence of these descriptions. We start with the calculation of \(j_{a,\pm}\).

\begin{itemize}
	\item[{\bf \S 1}] Let us first calculate the current associated to \( \delta_0 S =0\). In this case, we have 
	\be 
	\delta_0 B = 0\; ,
	\ee
	then
	\be 
	{\cal J}_{0, \pm} = \mathbb{J}_{0,\pm} = k_0^m\left( G_{mn} \mp B_{mn} \right) \partial_{\pm} X^n\; .
	\ee
	Therefore 
	\be 
	\begin{split} 
		{\cal J}_{0, +} & \equiv {\cal J}_0 =- \frac{u \partial v}{z^2}  + \frac{\partial z}{2 z} = \frac{- u \partial v + z\partial z}{z^2} - \frac{1}{2}\partial \ln z \\
		{\cal J}_{0, -} & \equiv \bar{{\cal J}}_0 =\frac{1}{2}\bar \partial \ln z\; .
	\end{split}
	\ee
	The important point is that these currents do not agree with the chiral currents we obtain in the WZNW model. But one can gauge the field \(B\) so that one of these currents agrees with the chiral current that one obtains from the WZNW model
	\be 
	\mathscr{J}_0 =\frac{1}{z^2}\left( z \partial z - u \partial v\right)\; .
	\ee
	In other words, we suppose \(B\to B+\dd \Lambda\), so that \(\mathscr{L}_a \Lambda = -J_a\). We find 
	\be 
	{\cal J}'_{0, \pm}= {\cal J}_{0, \pm} \pm j^\Lambda_{0, \pm}\; ,
	\ee
	and we conclude that 
	\be 
	j^\Lambda_{0,+}=\frac{1}{2}\partial\ln z\; , \quad 	j^\Lambda_{0,-}=\frac{1}{2}\bar \partial\ln z\; \Rightarrow \; J^\Lambda_0 = -\mathscr{L}_{k_0}\Lambda = \frac{1}{2}\dd \ln z\; ,
	\ee
	and from this expression, we notice that
	\be 
	\begin{split} 
		{\cal J}_{0, +} & \equiv  \mathscr{J}_0 \\
		{\cal J}_{0, -} & \equiv 0 \; .
	\end{split}
	\ee
	where we drop the prime.
	
	\item[{\bf \S 2}] Next, we want \(\delta_{+1} S=0\), where
	\be 
	\delta_{+1} B = 0\; ;
	\ee
	then
	\be 
	{\cal J}_{+1, \pm} = \mathbb{J}_{+1,\pm} = k_{+1}^m\left( G_{mn}\mp B_{mn} \right) \partial_{\pm} X^n\; .
	\ee
	Therefore
	\be 
	\begin{split} 
		{\cal J}_{+1, +} & \equiv {\cal J}_{+1} =- \frac{ \partial v}{z^2} \\
		{\cal J}_{+1, -} & \equiv \bar{{\cal J}}_{+1} =0 \; ,
	\end{split}
	\ee
	and this time, these objects are equal to the currents coming from the WZNW model. In this case we have
	\be 
	j^\Lambda_{+1,+}=0 \; , \quad 	j^\Lambda_{+1,-}=0\; \Rightarrow \; J_{+1} = -\mathscr{L}_{k_{+1}}\Lambda = 0\; .
	\ee
	
	\item[{\bf \S 3}] Finally, we can repeat the calculations for \(\delta_{-1} S=0\). Therefore 
	\be 
	\begin{split}
		\delta_{-1} B & =  \frac{\epsilon_{-1}}{z}\dd z \wedge \dd u = \epsilon_{-1}\left(\partial 	\ln z \bar{\partial} u - \bar{\partial}\ln z \partial u \right) \dd \sigma \wedge \dd \bar{\sigma}\\
		& = \left[ -(\bar{\partial} \epsilon_{-1}) u \partial \ln z + \bar{\partial}( \epsilon_{-1} u \partial \ln z) 
		+ (\partial \epsilon_{-1}) u \bar{\partial} \ln z -
		\partial( \epsilon_{-1} u \bar{\partial} \ln z)
		 \right]\dd \sigma \wedge \dd \bar{ \sigma} \; .
	\end{split}
	\ee
	The boundary term can be neglected and we conclude that
	\be 
	j_{-1, +} =   u \partial \ln z \; , \qquad j_{-1, -} =   u \bar{\partial} \ln z\; .
	\ee
	Given that \(\delta B\neq 0\), the currents
	\be 
	{\cal J}_{-1, \pm} = k_{-1}^m\left( G_{mn}\mp B_{mn} \right) \partial_{\pm} X^n\pm j_{-1,\pm}
	\ee
	give
	\be 
	\begin{split}
		{\cal J}_{-1, +} & = \frac{ u^2 \partial v}{z^2}-\frac{u \partial z	}{z} +  u \partial \ln z\\
		{\cal J}_{-1, -} &  =\bar{\partial} u - \frac{u \bar{\partial} z}{z}  -   u \bar{\partial} \ln z \; .
	\end{split}
	\ee
	
	Evidently we need to consider the gauge transformation \(B\to B+\dd \Lambda\) to obtain the chiral current 
	\be 
	\mathscr{J}_{-1} = \frac{u^2 \partial v}{z^2} - \frac{2 u \partial z}{z} + \partial u\; .	
	\ee
	Given that \(\mathscr{L}_{-1}\Lambda=-J_{-1}\), 
	\be 
	{\cal J}'_{-1,\pm} = {\cal J}_{0, \pm} \pm j^\Lambda_{0, \pm}\; ,
	\ee
	and from this expression we have 
	\be 
	j^\Lambda_{-1,+}=-2\frac{u\partial z}{z} + \partial u \; , \quad 	j^\Lambda_{-1,-}=-2\frac{u\bar{\partial} z}{z} + \bar{\partial} u\; .
	\ee
\end{itemize}

Using these expressions, we can find the gauge \(\Lambda\) and consequently the new B-field. Once we obtain the explicit expression for the B-field, the equivalence of the sigma model description of the AdS\(_3\) strings and the \(SL(2, \mathbb{R})\) is completely settled.


\bibliographystyle{utphys}
\bibliography{library.bib}
 
\end{document}